\documentclass[useAMS]{mn2e}
\usepackage{epsfig,natbib_vw,times}

\bibpunct[, ]{(}{)}{;}{a}{}{,}

\def \mnras {MNRAS}

\def \apjl {ApJL}

\def \be {\begin{equation}}
\def \ee {\end{equation}}

\def\gsim{\mathrel{\lower0.6ex\hbox{$\buildrel {\textstyle >}
 \over {\scriptstyle \sim}$}}}
\def\lsim{\mathrel{\lower0.6ex\hbox{$\buildrel {\textstyle <}
 \over {\scriptstyle \sim}$}}}

\def\m@th{\mathsurround=0pt }
\def\eqalign#1{\null\,\vcenter{\openup1\jot \m@th
 \ialign{\strut\hfil$\displaystyle{##}$&$\displaystyle{{}##}$\hfil
 \crcr#1\crcr}}\,}

\def \zabs {z_{\rm abs}}

\def \caii {Ca~{\sc ii}~}
\def \caiii {Ca~{\sc iii}~}

\def \mgii {Mg~{\sc ii}~}
\def \feii {Fe~{\sc ii}~}
\def \oii {[O~{\sc ii}]~}
\def \mgi {Mg~{\sc i}~}

\def \hi {H~{\sc i}~}
\def \nhi {$N$(H~{\sc i})~}

\def \EBV {\textit{E(B$-$V)}}

\title[Imaging of \caii Galaxies at
$z\sim1$]{$K$-band Imaging of strong Ca~{\sc ii}-absorber host galaxies at
$z\sim1$}
\author[P. Hewett \& V. Wild ]{Paul C. Hewett$^{1}$\thanks{phewett@ast.cam.ac.uk},
  and Vivienne Wild$^2$
\vspace*{6pt}\\
1. Institute of Astronomy, University of Cambridge, Cambridge CB3 0HA, UK \\
2. Max-Planck-Institut f\"{u}r Astrophysik, 85748 Garching, Germany\\}

\begin{document}

\maketitle
\begin{abstract}

We present $K$-band imaging of fields around 30 strong \caii absorption line
systems, at $0.7<z<1.1$, three of which are confirmed Damped Lyman-$\alpha$ systems.
A significant excess of galaxies is found within 6\farcs0 ($\simeq$50\,kpc) from
the absorber line-of-sight.  The excess galaxies are preferentially luminous
compared to the population of field galaxies.  A model in which field galaxies
possess a luminosity-dependent cross-section for \caii absorption of the form
$(L/L^*)^{0.7}$ reproduces the observations well.  The luminosity-dependent
cross-section for the \caii absorbers appears to be significantly stronger than the
established $(L/L^*)^{0.4}$ dependence for \mgii absorbers.  The associated galaxies
lie at large physical distances from the Ca~{\sc ii}-absorbing gas; we find a mean
impact parameter of 24\,kpc ($H_0=70\,{\rm km\,s}^{-1}\,{\rm Mpc}^{-1}$).
Combined with the observed number density of \caii absorbers the large physical
separations result in an inferred filling factor of only $\sim$10\, per cent.  The
physical origin of the strong \caii absorption remains unclear, possible
explanations vary from very extended disks of the luminous galaxies to associated
dwarf galaxy neighbours, remnants of outflows from the luminous galaxies, or tidal
debris from cannibalism of smaller galaxies.

\end{abstract}

\begin{keywords}
dust, extinction - galaxies: ISM, abundances - quasars: absorption lines

\end{keywords}

\section{Introduction}\label{sec:intro}


Quasar absorption line systems are caused by gas clouds lying along the
line-of-sight between a background quasar and the observer.  Visible through the
absorption of the background quasar light the systems provide a unique view of the
chemical composition of the gaseous component of the Universe over an extended
redshift range.  The strongest of the absorption systems probe the densest gaseous
regions associated with galaxies; possibly their enveloping gaseous haloes,
extended disks or gaseous outflows from winds.  Some absorption systems are
expected to probe the inter-stellar medium (ISM) of galaxies, allowing the study of the
build up of metals and dust within galaxies out to redshifts comparable to those of
the highest redshift quasars.  However, the precise link between the absorption--
and emission--selected galaxy population is currently unknown.


A key step towards gaining a complete understanding of the nature of quasar
absorption line systems is the compilation of a large sample in which the galaxies
associated with the absorption have been identified.  Such a sample would enable us
to link absorption properties with properties of the associated galaxies;
luminosity, morphological type and impact parameter for example.  A large sample is
necessary to understand the effects of factors such as the relative cross--section
of different morphological types, metallicity gradients within disks, internal dust
content and the effect of outflows on the observed properties of absorption line
systems.

Arguably one of the most interesting class of absorption line system for studying
galaxies and their immediate environments is damped Lyman-$\alpha$ (DLA) systems,
the absorbers with the highest neutral hydrogen column densities.  Unfortunately,
at $z\la1$, where deep imaging allows the identification of galaxies several
magnitudes below $L^*$, the Lyman-$\alpha$ line moves out of the observed frame
optical range and therefore the sample size of known DLAs over more than half the
age of the Universe is small.  While their low metallicities have led to the
suggestion that the host galaxies of DLAs may be low luminosity dwarf galaxies, the
small numbers of images obtained thus far have not established any significant
difference between the host galaxies of DLAs and the field galaxy population
\citep{2003ApJ...597..706C,2005ARA&A..43..861W}.


Recently \citet{2005MNRAS.361L..30W} showed that samples of the rare
\caii$\lambda\lambda3935,3968$ quasar absorption line systems could be compiled out
to cosmologically interesting redshifts, $z\simeq 1$, from quasar spectra in the
Sloan Digital Sky Survey (SDSS).  \caii is underabundant in the ISM of the Milky
Way due to both its ionisation level -- most calcium in the ISM of galaxies is
found as \caiii -- and its affinity for dust grains.  Strong \caii absorbers at
$0.8\la\zabs\la1.3$ cause significant reddening of the background quasar light by
dust, unlike DLAs and Mg~{\sc ii}--selected absorption line systems
\citep{2004MNRAS.354L..31M,2006MNRAS.367..211W,2006MNRAS.367..945Y}, and an
intriguing trend of increasing dust content with increasing \caii rest--frame
equivalent width ($W$) is observed.  Within the Milky Way,
\citet{2006MNRAS.367.1478H} suggest that \caii traces warm neutral gas clouds,
thus strong \caii lines at $z\sim1$ may probe the inner disks of chemically
evolved galaxies.  Alternatively, large \caii abundances may arise when dust
grains are destroyed, perhaps within shocks caused by major mergers
\citep{2005ApJ...622L.101W}.  The physics of dust is not well understood and
destruction of a small fraction of the grains could lead to a large relative
increase in gaseous \caii, consistent with the observed presence of dust in the
\caii systems.  A first step towards understanding the true nature of the \caii
systems is through imaging, which, for systems with redshifts below about unity, is
well within the capabilities of 4m-class telescopes.

Aside from the objective of understanding the origins of the
unusually strong \caii lines, \caii absorbers are expected to
contain large neutral hydrogen columns, with a significant number
above the nominal DLA limit \citep{2006MNRAS.367..211W}. With $\simeq$400
strong \caii absorbers in the SDSS Data Release 4 (DR4) quasar sample,
compared to the $\sim40$ known DLAs at $z\la1.65$
\citep{2006ApJ...636..610R}, they potentially represent an important
new method for the selection of galaxies by hydrogen cross section at
low and intermediate redshifts where the Lyman-$\alpha$ line is
currently observationally inaccessible.


In this paper we present the results of $K$-band imaging of the fields around 30
\caii absorbers with $0.7<z<1.2$.  The typical number density of $K$-band objects
is well known and low enough for single band imaging to be effective given the
expected small impact parameters between the absorbing galaxy and quasar.

The sample selection is described in Section \ref{sec:sample} and observations and
data reduction in Section \ref{sec:secobs}.  In Section \ref{sec:results} we
compare our $K$-band number counts to previous surveys and discuss the impact of
potential contamination from the targetting of quasar fields (quasar host and
companion galaxies).  Section \ref{sec:props} presents our results for the
luminosity, impact parameters and morphology of the \caii absorber galaxies.  In
Section \ref{sec:disc} we first discuss these results in comparison to strong \mgii
absorbers and DLAs, then we consider the implications for the present day star
formation rate derived in \citet{2007MNRAS.374..292W}.  Finally, in Section
\ref{sec:concl}, we discuss viable models for the true nature of \caii absorption
line systems.

Optical apparent and absolute magnitudes taken from the SDSS catalogues are left in
the AB-system, while the observed $K$-band magnitudes are given on the Vega system,
as employed by 2MASS.  Conversion between the systems can be achieved using the
relations $i{\rm (AB)}=i{\rm (Vega)}+0.37$, $K_{\rm 2MASS}{\rm (AB)}= K_{\rm
2MASS}{\rm (Vega)}+1.86$ and, similarly, for the Mauna Kea Observatory $K$-band,
$K_{\rm MKO}{\rm (AB)}=K_{\rm MKO}{\rm (Vega)}+1.90$ \citep[][with $m_V$=+0.03 for
Vega]{2006MNRAS.367..454H}.  We use a flat cosmology with $\Omega_\Lambda=0.7$,
$\Omega_M=0.3$, $H_0=100\,h\,{\rm km\,s}^{-1}\,{\rm Mpc}^{-1}$ and $h=0.7$ throughout the
paper.

\section{Sample Selection}\label{sec:sample}

The absorber sample of 30 objects to be imaged was selected from the catalogue of
345 \caii absorbers described in \citet{2007MNRAS.374..292W}.  The absorber
redshift range $0.8 \la z_{abs} \la 1.1$ was chosen to probe the old stellar
populations associated with absorbers at a significant lookback-time.  Three
absorbers in the catalogue with \nhi measurements from \citet{2006ApJ...636..610R}
were preferentially targeted, including two with $0.7 < z_{abs} < 0.8$.  To avoid
complications resulting from the detection of galaxies associated with the
background quasar, targets with $z_{quasar} \ga z_{abs}+0.2$ were selected, with
the majority of targets possessing differences of $\Delta z > 0.5$.  Otherwise, the
make up of the imaging sample was determined by observing constraints, including
accessibility of the fields from the United Kingdom Infra-Red Telescope (UKIRT) and
the availability of a suitable guide star.  The sample includes the full range of
quasar brightness and \caii equivalent width within the catalogue of 345 absorbers.

To summarise, the 30 \caii absorbers were selected according to the 
following criteria:

\begin{itemize}
\item $0.8 < z_{abs} < 1.2$, {\it or}, $0.7 < z_{abs} \le 0.8$ with \nhi from 
\citep{2006ApJ...636..610R}
\item $z_{quasar} > z_{abs}+0.2$
\item $7^h 30^m < \alpha < 18^h 00^m$ and $-1^\circ < \delta < 58^\circ$ 
\item Suitable guide star available
\end{itemize}

In addition, a small sample of five `control' quasars were imaged. The quasars
were selected from \citep{2005AJ....130..367S} to have $z_{quasar} \simeq 1.5$,
$m_i \simeq 18.0$ and to possess no detected absorption systems with $z_{abs}
< 1.2$. Finally, a single `blank field' was also imaged. Table 1 includes the
observing log for the target and control sample, along with the positions and 
redshifts of the targets.

\begin{table*}
  \begin{center}
    \caption{\label{tab:obs} Journal of Observations
    }
\begin{minipage}{17.5cm}

\begin{tabular}{lccccccccc} \hline
Name & $z_{abs}$ & $z_{quasar}$ & $m_{i,quasar}$ & $m_{K,quasar}$ & Exposure & Seeing &
$\mu_{K}$ \footnote{The surface brightness threshold used to define the galaxy
catalogue in the $K$-band image (Section 3.4)}&  
PSF \footnote{The source of the
PSF-template employed during subtraction of the quasar PSF. F: the
template PSF was created from a bright star(s) within the frame; L: the
template PSF was selected from a library of PSFs created from
specially targetted stars or stars in other quasar frames.}  &Observing
Date \\
                         &       &       &       &      & (s)  & (")  &(mag\,arcsec$^{-2}$) &  & UT \\ \hline
SDSS J074804.06+434138.5 & 0.898 & 1.836 & 18.44 & 16.7 & 2520 & 0.64 & 21.1 & F & 2006-Apr-17 \\
SDSS J080735.97+304743.8 & 0.969 & 1.255 & 18.53 & 16.4 & 1800 & 0.84 & 20.9 & F & 2006-Apr-11 \\
SDSS J081930.35+480825.8 & 0.903 & 1.994 & 17.64 & 16.5 & 2520 & 0.64 & 21.2 & L & 2006-Apr-16 \\
SDSS J083819.85+073915.1 & 1.133 & 1.568 & 18.46 & 16.7 & 2520 & 0.62 & 21.2 & F & 2006-Apr-15 \\
SDSS J085221.25+563957.6 & 0.844 & 1.449 & 18.58 & 16.2 & 2520 & 0.56 & 21.0 & F & 2006-Apr-16 \\
SDSS J085556.63+383232.7 & 0.852 & 2.065 & 17.57 & 15.5 & 2520 & 0.73 & 21.2 & F & 2006-Apr-14 \\
SDSS J093738.04+562838.8 & 0.978 & 1.798 & 18.49 & 16.2 & 2520 & 0.69 & 21.1 & F & 2006-Apr-14 \\
SDSS J095307.05+111140.8 & 0.980 & 1.877 & 18.79 & 16.6 & 2520 & 0.75 & 21.1 & L & 2006-Apr-15 \\
SDSS J095352.69+080103.6 & 1.023 & 1.720 & 17.40 & 15.6 & 2520 & 0.66 & 21.1 & F & 2006-Apr-14 \\
SDSS J100339.44+101936.8 & 0.816 & 1.844 & 18.36 & 16.7 & 2520 & 0.66 & 21.1 & F & 2006-Apr-17 \\
SDSS J105930.50+120532.8 & 1.050 & 1.570 & 18.84 & 16.5 & 2520 & 0.53 & 21.1 & L & 2006-Apr-15 \\
SDSS J110729.03+004811.2 & 0.740 & 1.391 & 17.10 & 15.2 & 2520 & 0.47 & 21.1 & F & 2006-Apr-16 \\
SDSS J113357.55+510844.9 & 1.029 & 1.576 & 18.28 & 16.7 & 2520 & 0.82 & 21.2 & F & 2006-Apr-14 \\
SDSS J115523.97+480141.6 & 1.114 & 1.542 & 18.82 & 16.7 & 2520 & 0.51 & 21.1 & L & 2006-Apr-16 \\
SDSS J120300.96+063440.8 & 0.862 & 2.182 & 18.47 & 15.6 & 2520 & 0.71 & 21.2 & L & 2006-Apr-14 \\
SDSS J122144.62-001141.8 & 0.929 & 1.750 & 18.52 & 16.9 & 2520 & 0.73 & 21.1 & F & 2006-Apr-14 \\
SDSS J122756.35+425632.4 & 1.045 & 1.310 & 17.17 & 15.4 & 2520 & 0.56 & 21.2 & F & 2006-Apr-15 \\
SDSS J124659.81+030307.6 & 0.939 & 1.178 & 18.81 & 16.6 & 2520 & 0.64 & 21.2 & F & 2006-Apr-11 \\
SDSS J130841.19+133130.5 & 0.951 & 1.954 & 18.60 & 16.6 & 2520 & 0.60 & 21.2 & F & 2006-Apr-11 \\
SDSS J132323.78-002155.2 & 0.716 & 1.388 & 17.61 & 15.4 & 2520 & 0.64 & 21.2 & L & 2006-Apr-17 \\
SDSS J132346.05-001819.8 & 1.087 & 1.842 & 18.84 & 16.4 & 2520 & 0.46 & 21.1 & F & 2006-Apr-16 \\
SDSS J140444.19+551636.9 & 1.070 & 1.589 & 18.46 & 16.5 & 2520 & 0.62 & 21.1 & F & 2006-Apr-14 \\
SDSS J145633.08+544831.6 & 0.879 & 1.518 & 17.94 & 16.3 & 2520 & 0.56 & 21.1 & L & 2006-Apr-11 \\
SDSS J151247.48+573843.4 & 1.044 & 2.135 & 18.69 & 15.7 & 2520 & 0.62 & 21.1 & L & 2006-Apr-11 \\
SDSS J153503.36+311832.4 & 0.904 & 1.510 & 17.79 & 15.8 & 2520 & 0.62 & 21.1 & L & 2006-Apr-14 \\
SDSS J155529.40+493154.9 & 0.893 & 1.831 & 17.64 & 15.6 & 2520 & 0.51 & 21.1 & L & 2006-Apr-14 \\
SDSS J160932.95+462613.3 & 0.966 & 2.361 & 18.67 & 16.6 & 3240 & 0.56 & 21.2 & L & 2006-Apr-11 \\
SDSS J164350.94+253208.8 & 0.885 & 1.560 & 17.70 & 16.0 & 2520 & 0.47 & 21.0 & F & 2006-Apr-17 \\
SDSS J172739.01+530229.2 & 0.945 & 1.442 & 17.97 & 16.1 & 2520 & 0.47 & 21.0 & F & 2006-Apr-16 \\
SDSS J173559.95+573105.9 & 0.872 & 1.824 & 18.18 & 16.5 & 2160 & 0.51 & 20.9 & F & 2006-Apr-17 \\
SDSS J093411.14+000519.7 & --- & 1.534 & 18.08 & 16.5 & 2520 & 0.66 & 21.0 & F & 2006-Apr-17 \\
SDSS J114445.93+055053.8 & --- & 1.596 & 18.02 & 16.5 & 2520 & 0.69 & 21.1 & L & 2006-Apr-17 \\
SDSS J142519.70+012043.2 & --- & 1.599 & 18.09 & 15.9 & 2520 & 0.42 & 21.1 & F & 2006-Apr-16 \\
SDSS J150306.34+011345.8 & --- & 1.521 & 18.09 & 16.0 & 2520 & 0.53 & 21.1 & F & 2006-Apr-17 \\
SDSS J160108.43+422937.5 & --- & 1.531 & 18.05 & 16.2 & 2520 & 0.51 & 21.1 & F & 2006-Apr-17 \\
BLANK J132224.4-002155 & --- & --- & --- & --- & 1360 & 0.62 & 20.9 & --- & 2006-Apr-16 \\
\hline

\end{tabular}
\vspace*{-0.4cm}
\end{minipage}
\end{center}
\end{table*}      

\section{Observations and data reduction}\label{sec:secobs}

The key aim of our observation and data reduction strategy was to
obtain a well characterised point spread function (PSF) to allow
accurate subtraction of the quasar image and recover candidate
absorber host galaxies as faint and close to the quasar as possible.

\subsection{Observations}\label{sec:obs}

Imaging was undertaken on the 3.8m UKIRT using the Fast-Track Imager (UFTI)
near-infared imager \citep{2003SPIE.4841..901R} and a Mauna Kea Observatory
$K$-band filter, over the nights of 2006 April 11 and 14-17 UT.  The
transparency on 2006 April 11 was variable, but otherwise transparency was good
and conditions photometric for a significant fraction of the observations.  The
median on-chip seeing was 0\farcs62 full width at half maximum (FWHM), with a
full range of 0\farcs40--0\farcs85 FWHM, which was well sampled by the UFTI
image scale of 0\farcs091 per pixel.  The on-chip seeing generally improved over
the first few hours of observation each night.

An isolated $K_{\rm 2MASS}\simeq 12.3$ star was observed before and after each
target to provide a reference PSF for use in the subsequent quasar-subtractions.
The default observing sequence for a target consisted of a single 9-point dither of
the PSF-star with 5\,s exposures, followed by a series of seven 9-point dithers for
the target with 40\,s exposures, finally, the single 9-point dither of the PSF-star
was repeated.  The dither offset distance employed was 20\arcsec--22\arcsec for
both PSF-star and target observations (the small variation from step-to-step was
designed to avoid the repeated use of identical columns on the detector).  The
sequence thus resulted in two 90\,s exposures of the PSF-star and a single 2520\,s
exposure for the target.  For a small number of targets, the number of 9-point
dithers was varied in order to accommodate observing constraints.  Table
\ref{tab:obs} presents the journal of observations and includes the exposure time
(Col.  5) and on-chip seeing (Col.  6) obtained for each target.  Appropriate dark
frames were obtained for each PSF-target-PSF sequence.

Photometric zero-points for the target images were obtained by
adopting $K_{\rm MKO}=K_{\rm 2MASS}$ for the PSF-stars and applying
the zero-points so determined to the target frames.  The two PSF-star
derived zeropoints for each target were consistent to better than
0.05\,mag for all but the observations made on 2006 April 11.  The
zero-points for the first night indicated attenuation by cloud of
0.1-0.3\,mag for several of the observations, but, conditions were
changing only slowly and the consistency between pre- and post-target
PSF-observations was at worst 0.09\,mag.  The accuracy of the
magnitude zero-points for the target observations are thus
$\la$0.05\,mag.

\subsection{Data reduction}

Data reduction techniques employed standard routines available in {\sc
iraf}\footnote{{\sc iraf} is distributed by the National Optical
Astronomy Observatories, which are operated by the Association of
Universities for Research in Astronomy, Inc.  under cooperative
agreement with the National Science Foundation.}  The frames were
first dark-subtracted using the appropriate calibration frames.

Flatfielding of each PSF-star and target dither was performed using a
three-stage procedure.  First, a crude flatfield was generated using a
1.75$\sigma$-clipped average of all the PSF-star and target frames
taken on a given night.  The resulting nightly flatfields were
normalised and divided into the individual frames.  The {\sc objmasks}
routine was then used to exclude areas of the frames affected by
astronomical objects and the generation of the nightly average
flatfield repeated.  Division of the individual frames by these
improved flatfields produced good results for both PSF-star and target
frames, but low-level structure in the sky was still visible at the 1
per cent level.  The final step in the flatfielding involved rerunning
the {\sc objmasks} routine on the flatfielded frames and then
generating a 2$\sigma$-clipped average of the 9 (63) frames in each
PSF-star (target) sequence. Division of each frame by the resulting
final flatfield reduced the structure in the sky to $\la 0.3$ per
cent.

The master PSF-star (target) image was generated from a
5$\sigma$-clipped average of the 9 (63) frames, employing the
appropriate XY-offsets for the dither pattern. A bad-pixel mask,
derived from the dark frames and the variance of the nightly flatfield
frames, was employed to exclude bad pixels from the averaging. The
resulting frames were both cosmetically clean and showed no detectable
structure in the sky. The combined image was then clipped to retain
just the central fully-exposed $536 \times 536$ pixel ($49\arcsec
\times 49\arcsec$) region.

\subsection{Point spread function subtraction}\label{sec:psfsub}

\begin{figure*}

  \includegraphics[scale=0.95]{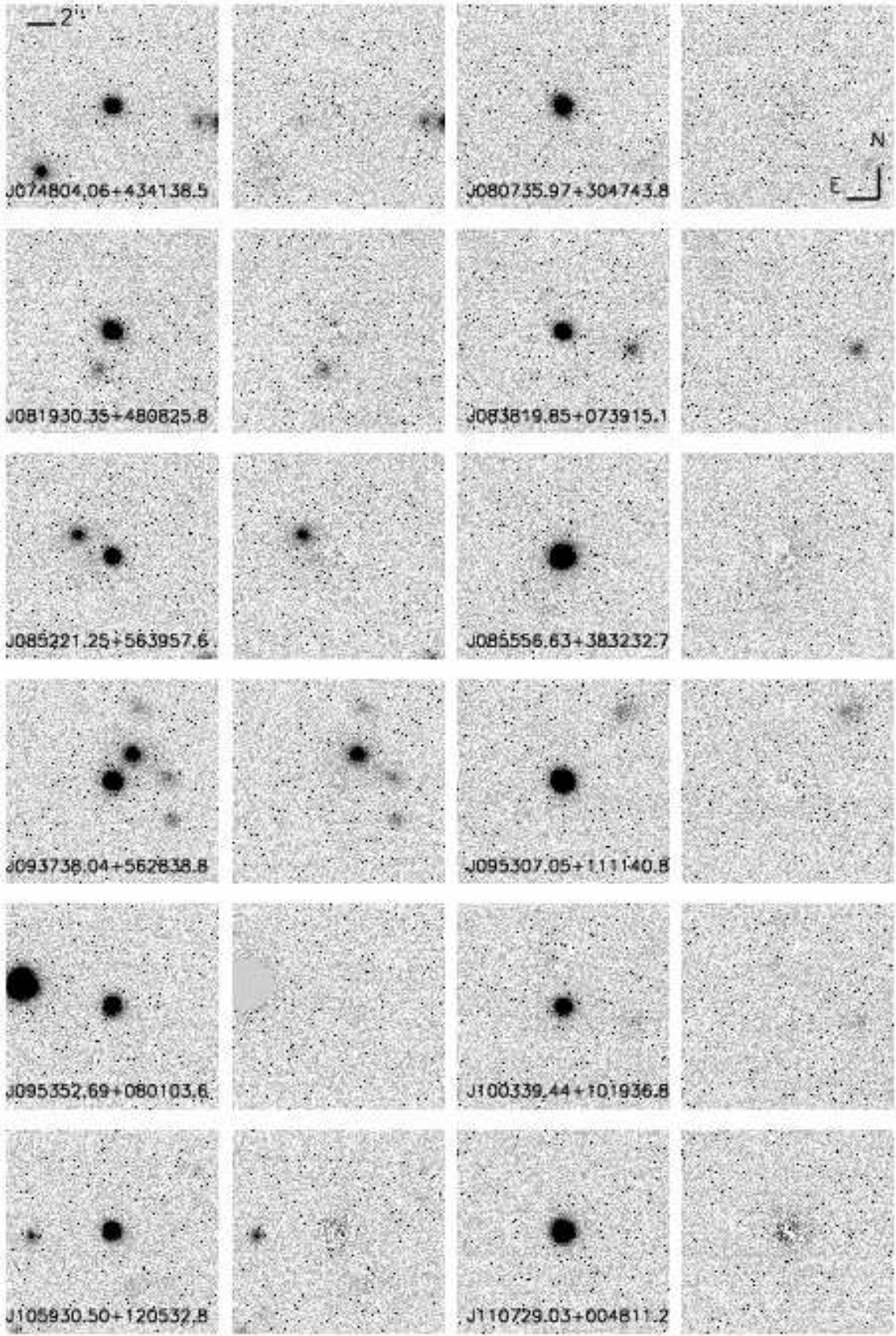}

  \caption{Images of the 30 \caii-absorber quasars and the five control
quasars. The left-hand panel in each pair shows the $K$-band image of
the quasar field, while the right-hand panel shows the quasar field after
subtraction of the quasar. Each images is 15\farcs \ on a side. A scale-bar is
shown in the image at top left on each page and the orientation of the images
is indicated in the image at top right on each page.}
\label{fig:tile} 
\end{figure*}
\begin{figure*}
  \includegraphics[scale=0.95]{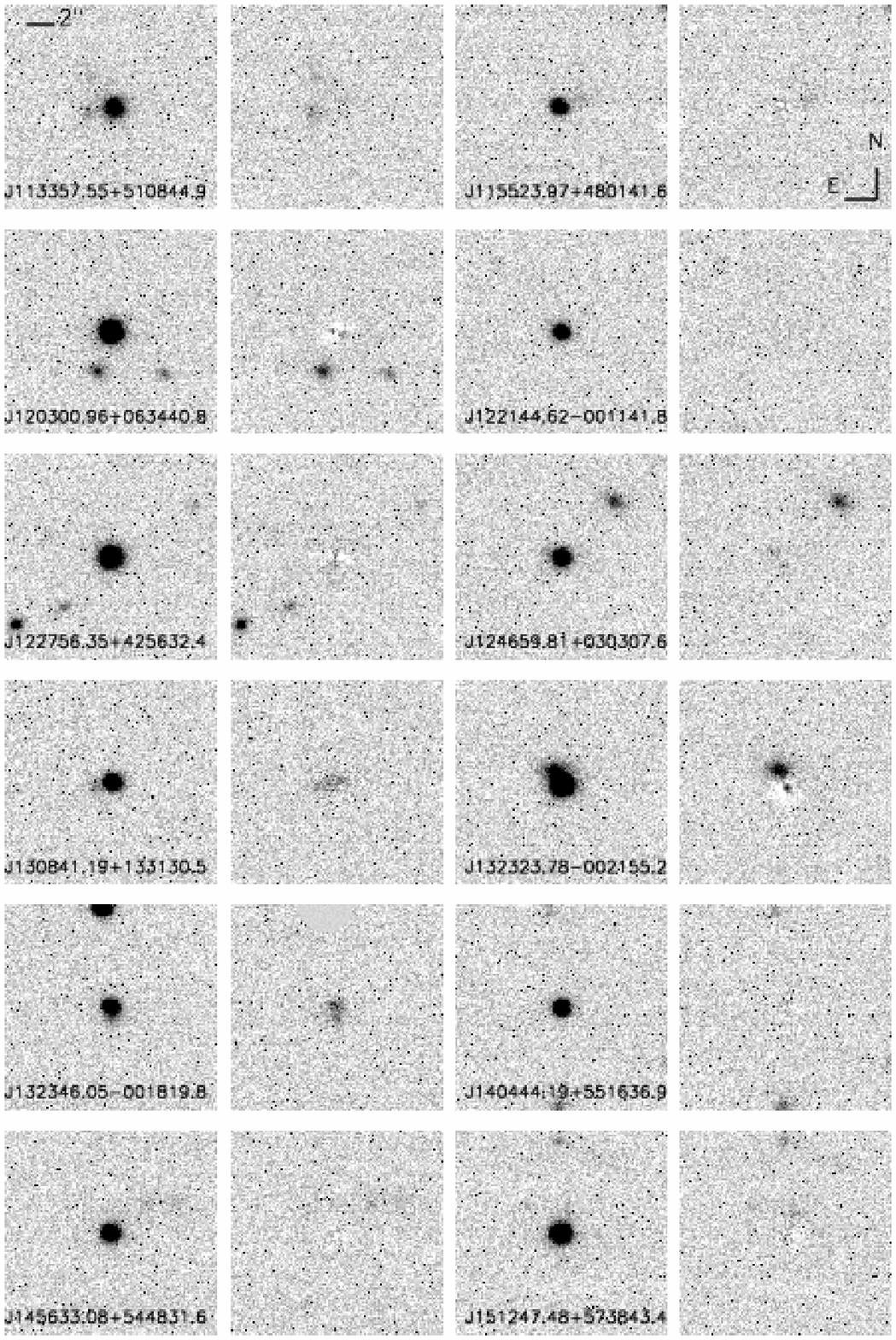}
\begin{center} Figure~\ref{fig:tile} (continued)\end{center}
\end{figure*}
\begin{figure*}
  \includegraphics[scale=0.95]{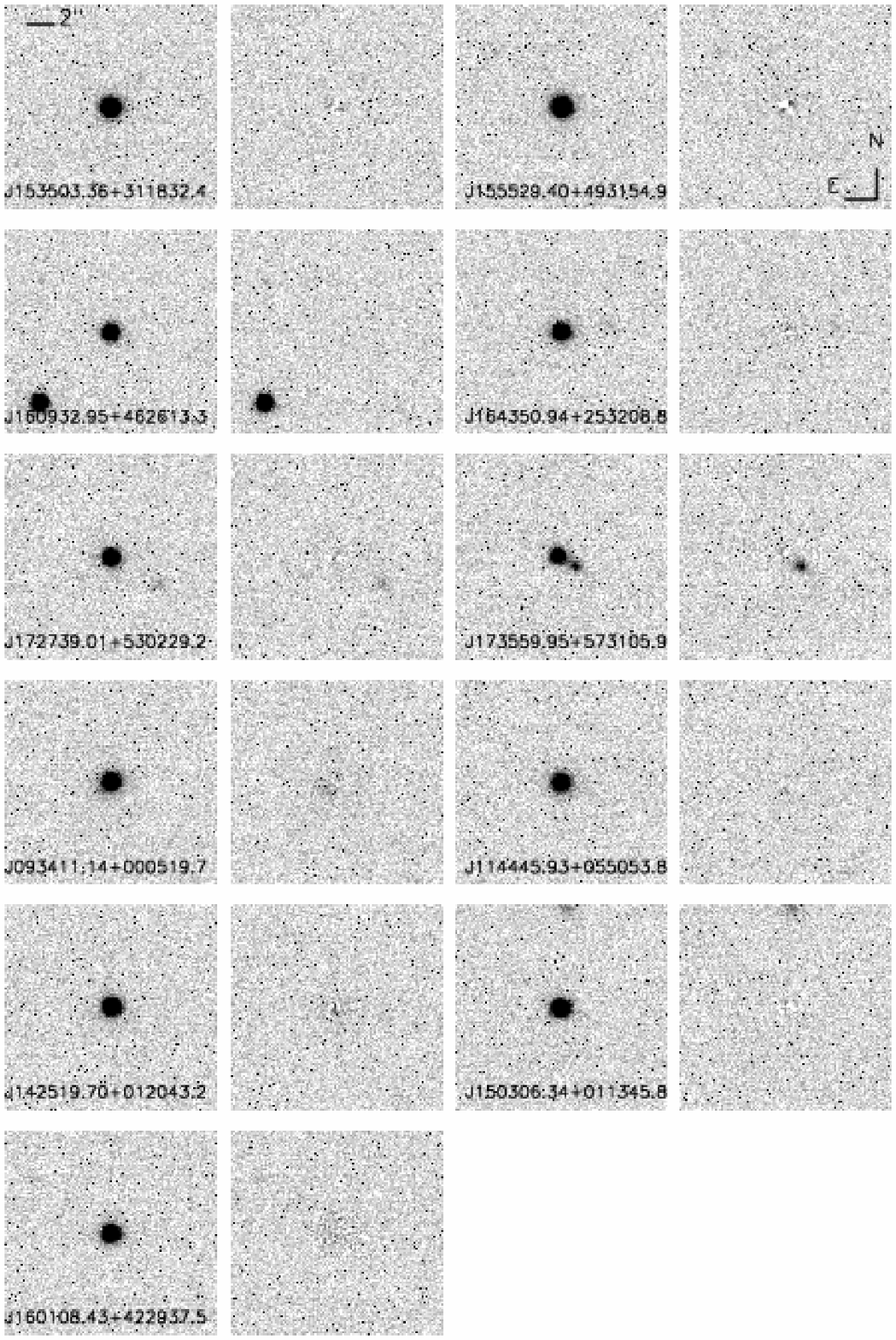}
\begin{center} Figure~\ref{fig:tile} (continued)\end{center}
\end{figure*}

In order to identify faint galaxies with small projected separations
from the quasars, subtraction of the quasar images was undertaken
using the standard sequence of {\sc daophot} routines
\citep{1987PASP...99..191S} in {\sc iraf}.  A 21 pixel (1\farcs9)
radius aperture was used to define the PSFs and also for the quasar
image subtractions.  The PSF-star images were analysed using the {\sc
psf} routine to produce a library of template-PSFs.  All the PSF-star
images were best-fit by a Moffat profile, usually with $\beta=1.5$,
although the images obtained in the poorest seeing typically produced
a best-fit with $\beta=2.5$.

The target frames were then inspected visually and additional template-PSFs were
constructed using the {\sc daophot} routines, where a single stellar image
brighter than $m_{quasar}+0.5$ was present in the frame.  The number of target
frames for which two or more such suitable stars were present was small but in
such cases a single template-PSF was constructed.  Care was taken to ensure that
none of the stellar images possessed peak counts taking them into the non-linear
exposure--count regime of UFTI.

The subtraction of the quasar images was undertaken using the {\sc substar}
routine.  As expected, the template-PSFs derived from a particular target frame
produced the most satisfactory subtractions for the target quasar.  Where such a
template-PSF was not available, a suitable template-PSF was chosen from the full
library based on the seeing, ellipticity and orientations of fainter stellar
images in the target frame.  Table \ref{tab:obs}, Col.  9 indicates whether the
adopted PSF-subtraction employed a library (L) or target frame (F) PSF.

The results of all the subtractions were examined visually and a series of tests
carried out to verify the reliability of the quasar-image removal.  In cases
where a faint image was revealed by the subtraction, other stellar images (when
present) in the target frame were also processed to ensure that similar `images'
were not revealed.  Similarly, when low-level systematic residuals, usually of
the `butterfly' or `clover-leaf' type, were present\footnote{see the subtracted
images of J120300.96+063440.8 and J155529.40+493154.9 respectively, in Fig.  1
for clear examples of the residual morphologies}, other stellar images were
processed to ascertain that the same form of residual persisted.  In all cases,
irrespective of whether faint images or residuals were apparent, a number of
different PSF-templates were tested to check that the results of the image
subtraction was not strongly dependent on the exact PSF-template used.  Fig.  1
shows the pre- and best post-subtracted quasar images for the 30 \caii quasars
and the five control quasars.  Each image pair shows a $15\farcs0 \times
15\farcs0$ region centred on the quasar, with the pre-subtraction image
presented on the right-hand side.  A scale-bar is shown in the top right-hand
panel of each page.

The combination of the UFTI instrument, with its small 0\farcs091 pixels, and an
observing strategy in which high signal-to-noise ratio target and PSF-star
observations were obtained was designed to maximise the accuracy of the PSF
characterisation.  The resulting image subtractions were well-behaved with the
form of the residuals showing consistent behaviour among multiple stars within
the same frame, and as a function of the degree of similarity between the
PSF-templates and the image profiles.  However, the magnitude range of the
quasars, $15.4 \le K \le 16.9$, causes a significant variation to exist in the
magnitude of the faintest galaxies that may be detected reliably, as the
PSF-subtraction is limited in accuracy by a constant percentage of the quasar
flux.

To determine the effectiveness of the PSF-subtraction a series of simulations
were undertaken using the {\sc artdata} package within {\sc iraf}.  Synthetic
galaxies, modelled pessimistically as pure exponential disks, were blurred to
match the seeing using a Moffat profile.  To match the appearance of galaxies in
the images, these synthetic galaxies were constructed with intrinsic half-light
radii of 0\farcs1 to 0\farcs25, corresponding to spatial half-light diameters of
$\sim$2-5\,kpc, and a range of apparent magnitude $19.0 \le K \le 20.0$.  The
synthetic galaxies were then added to the actual frames of absorber and control
quasars that were not found to possess close companion galaxies.  Three
quasar-galaxy separations of $\Delta\theta=0\farcs0, 0\farcs5$ and $1\farcs0$
were explored, using quasar images covering the full magnitude range of quasars
present in the sample.

The results of the simulations were encouraging.  At separations
$\Delta\theta \ge 1\farcs0$ the galaxy images are visible in the
pre-PSF-subtracted images for all but the brightest quasars, and all
galaxies as faint as $K=20.0$ are recovered. Down to separations of
$\Delta\theta \ge 0\farcs5$ all galaxies with $K=19.5$ are revealed
via the PSF-subtraction.  Fainter than $K=19.5$, galaxies can be
detected down to $K=20.0$ only for the faintest third of the quasar
images ($K \ge 16.5$).  The recovery rate for essentially
zero-separations, $\Delta\theta \la 0\farcs25$, is poor.  Galaxies
with $K \ge 19.5$ cannot be recovered for any of the quasars and
galaxies with $19.0 \le K < 19.5$ can only be recovered for quasars
with $K \ge 16.0$.  The rapid degredation in the ability to
recover faint galaxies at the very smallest separations comes about
primarily because of the presence of the characteristic `butterfly' or
`clover-leaf' residuals, which are caused by very small differences in
the PSFs and potential confusion from the cores of quasar host
galaxies.  In summary, the PSF-subtraction is capable of recovering
galaxies with percentage fluxes of the quasar of 1, 1.5 and 7 per cent
at separations of $\Delta\theta=1\farcs0, 0\farcs5$, and $0\farcs0$
respectively. Notes regarding the results of the PSF-subtraction for a
small number of quasars are included in Appendix \ref{sec:app}.

\subsection{Object catalogue generation}

The high-resolution of the raw images, necessary for the image
subtractions, was not optimal for the detection of faint low-surface
brightness images within the target exposures.  Image catalogues,
including the detection of the target quasar or
quasar+close\_companion, were constructed by applying {\sc SExtractor}
\citep{1996A&AS..117..393B} to the pre- and post-PSF-subtracted
images, rebinned by a factor of two to produce 0\farcs182 pixels.  A
1$\sigma$ threshold above the sky level, equivalent to a
surface-brightness threshold $\mu_K \simeq 21.1\,$mag\,arcsec$^{-2}$,
was employed with a Gaussian detection kernel matched to the seeing of
the target frame.  Cols.  6 and 7 of Table \ref{tab:obs} list the
seeing and surface-brightness thresholds for each target frame.
Visual inspection of the target frames confirmed the effectiveness of
the {\sc SExtractor} analysis and the resultant objects catalogues
were not sensitive to small changes in the detection parameters.

All the detected objects in the frames were inspected visually and a
small number ($\simeq$10) at the extreme edges of the frames, or
resulting from a faint satellite trail in one target frame, were
eliminated.  The number of objects detected in each target frame
(typically $\simeq$10) was too small for the morphological image
classification provided by {\sc SExtractor} to prove reliable.
Consequently, all objects with $K \le$18.5 were classified visually.
PSF-fitting was employed in the few cases where any doubt concerning
the classification arose.

The {\sc SExtractor} {\sc mag\_best} parameter was used to provide the
magnitude estimate for the detected objects.  The magnitudes of the
quasars were taken from the values provided by the {\sc daophot} {\sc
substar} routine.  After applying the zero-points for each target
frame (Section \ref{sec:obs}) we detect $K=20.0$ objects in all
frames, with the catalogues from the longest exposures in the best
seeing containing objects as faint as $K=21.0$.  Magnitude errors
are estimated from the detected electrons per object and an additional
approximation, to take account of the uncertainty in the
sky-background determination, calculated by shifting the sky-level
under the total area of the image by 20\,per cent of the $1\sigma$
pixel-to-pixel fluctuations in the sky background.

Table \ref{tab:galcat} presents the catalogue of nearest neighbour
galaxies, including {\it all} galaxies within 6\farcs0 of each quasar.
Cols.  2, 3, and 4 give the quasar-galaxy separations in arcseconds
and Col.  5 gives the galaxy $K$-band magnitude.  The final column
specifies whether the galaxy is the first-, second-, third- or
fourth-nearest neighbour to the quasar.

\begin{table*}
  \begin{center}
\begin{minipage}{14.5cm}
    \caption{\label{tab:galcat} All companion galaxies within 6\farcs0
    from the absorber, or the nearest neighbour if no closer
    galaxies are seen. Col 2--4: angular offsets from the
    background quasar, Col 5: apparent magnitudes, Col 6: first,
    second, third or fourth nearest neighbour, Col 7-8: probability of
    being a field galaxy, Col 9: Absolute magnitude.
    }

\begin{tabular}{ccccccccc} \hline
Quasar Name & $\Delta$RA & $\Delta$Dec & $\Delta\theta$ & $K$ &
Neighbour & $P_{\rm Field}(m)$\footnote{Probability the galaxy is a
field galaxy, given its apparent magnitude.} & $P_{\rm
Field}(\Delta\theta)$\footnote{Probability the galaxy is a field
galaxy, given its projected distance from the absorber.}  &
$M_K$\footnote{Vega magnitudes, H$_0= 100\,{\rm km\,s^{-1}Mpc^{-1}}$}  \\

            &     (")    &     (")     &  (")       &       &           &  &  &       \\ \hline
\multicolumn{9}{l}{Absorber Quasars}\\
J074804.06+434138.5 & 2.7 & -1.2 & 2.9 & 19.26$\pm$0.21 & 1 & 0.040 & 0.267 & -23.02 \\
J080735.97+304743.8 & -5.8 & -4.2 & 7.2 & 19.31$\pm$0.12 & 1 & ---  & ---   & ---    \\
J081930.35+480825.8 & 0.9 & -2.8 & 2.9 & 18.53$\pm$0.18 & 1 & 0.291 & 0.267 & -23.76 \\
J083819.85+073915.1 & -4.9 & -1.3 & 5.0 & 18.40$\pm$0.14 & 1 & 0.323 & 0.419 & -24.43 \\
J085221.25+563957.6 & 1.7 & -0.2 & 1.7 & 18.82$\pm$0.19 & 1 & 0.396 & 0.133 & -23.31 \\
J085221.25+563957.6 & 2.4 & 1.5 & 2.8 & 17.18$\pm$0.14 & 2 & 0.182 & 0.267 & -24.94 \\
J085556.63+383232.7 & 1.1 & -3.9 & 4.0 & 19.35$\pm$0.11 & 1 & 0.246 & 0.480 & -22.80 \\
J093738.04+562838.8 & -1.4 & 1.9 & 2.4 & 16.80$\pm$0.11 & 1 & 0.389 & 0.267 & -25.68 \\
J093738.04+562838.8 & -3.8 & 0.3 & 3.8 & 18.00$\pm$0.18 & 2 & 0.107 & 0.467 & -24.48 \\
J093738.04+562838.8 & -4.2 & -2.8 & 5.0 & 18.90$\pm$0.19 & 3 & 0.793 & 0.419 & -23.58 \\
J093738.04+562838.8 & -1.9 & 5.2 & 5.6 & 18.67$\pm$0.20 & 4 & 0.368 & 0.419 & -23.81 \\
J095307.05+111140.8 & -8.0 & -3.7 & 8.8 & 19.28$\pm$0.19 & 1 & ---  & ---   & ---    \\
J095352.69+080103.6 & -4.4 & 5.1 & 6.7 & 18.19$\pm$0.16 & 1 & ---  & ---   & ---    \\
J100339.44+101936.8 & -5.0 & -1.1 & 5.1 & 19.52$\pm$0.18 & 1 & 0.328 & 0.419 & -22.53 \\
J105930.50+120532.8 & -1.6 & -4.1 & 4.4 & 19.39$\pm$0.14 & 1 & 0.116 & 0.480 & -23.25 \\
J105930.50+120532.8 & 5.7 & -0.4 & 5.7 & 18.60$\pm$0.13 & 2 & 0.221 & 0.419 & -24.05 \\
J110729.03+004811.2 & 16.1 & 6.3 & 17.3 & 18.98$\pm$0.19 & 1 & ---  & ---   & ---    \\
J113357.55+510844.9 & 1.6 & -0.3 & 1.6 & 18.06$\pm$0.12 & 1 & 0.170 & 0.133 & -24.53 \\
J115523.97+480141.6 & -1.6 & 0.6 & 1.7 & 19.28$\pm$0.17 & 1 & 0.219 & 0.133 & -23.51 \\
J120300.96+063440.8 & 0.9 & -2.9 & 3.0 & 18.16$\pm$0.14 & 1 & 0.248 & 0.267 & -24.02 \\
J120300.96+063440.8 & -3.7 & -3.0 & 4.8 & 19.26$\pm$0.14 & 2 & 0.069 & 0.480 & -22.92 \\
J122144.62-001141.8 & -2.3 & -9.0 & 9.3 & 19.28$\pm$0.14 & 1 & ---  & ---   & ---    \\
J122756.35+425632.4 & 4.3 & 1.8 & 4.6 & 19.95$\pm$0.18 & 1 & 0.905 & 0.480 & -22.69 \\
J122756.35+425632.4 & 3.3 & -3.7 & 4.9 & 19.21$\pm$0.17 & 2 & 0.839 & 0.480 & -23.42 \\
J124659.81+030307.6 & 0.8 & 0.2 & 0.8 & 19.24$\pm$0.17 & 1 & 0.110 & 0.089 & -23.14\\
J124659.81+030307.6 & -3.7 & 4.1 & 5.6 & 17.96$\pm$0.15 & 2 & 0.066 & 0.419 & -24.42 \\
J130841.19+133130.5 & 0.5 & -0.0 & 0.5 & 18.30$\pm$0.14 & 1 & 0.166 & 0.089 & -24.11 \\
J132323.78-002155.2 & 0.4 & 1.0 & 1.1 & 17.26$\pm$0.13 & 1 & 0.416 & 0.133 & -24.48 \\
J132346.05-001819.8 & 0.0 & -0.3 & 0.3 & 17.93$\pm$0.15 & 1 & 0.507 & 0.089 & -24.80 \\
J140444.19+551636.9 & 0.8 & 7.0 & 7.1 & 19.46$\pm$0.18 & 1 & ---  & ---   & ---    \\
J145633.08+544831.6 & -2.6 & 2.2 & 3.4 & 19.42$\pm$0.21 & 1 & 0.319 & 0.467 & -22.80 \\
J145633.08+544831.6 & -4.7 & 2.3 & 5.2 & 19.55$\pm$0.16 & 2 & 0.223 & 0.419 & -22.67 \\
J151247.48+573843.4 & -0.6 & 1.5 & 1.6 & 19.61$\pm$0.27 & 1 & 1.000 & 0.133 & -23.03 \\
J153503.36+311832.4 & -17.2 & -3.5 & 17.6 & 19.22$\pm$0.21 & 1 & ---  & ---   & ---    \\
J155529.40+493154.9 & -18.0 & 17.8 & 25.3 & 17.61$\pm$0.15 & 1 & ---  & ---   & ---    \\
J160932.95+462613.3 & -2.9 & -11.5 & 11.8 & 19.76$\pm$0.22 & 1 & ---  & ---   & ---    \\
J164350.94+253208.8 & -3.4 & 0.4 & 3.5 & 19.96$\pm$0.23 & 1 & 0.237 & 0.467 & -22.28 \\
J172739.01+530229.2 & -3.3 & -1.8 & 3.7 & 19.38$\pm$0.19 & 1 & 0.124 & 0.467 & -23.02 \\
J173559.95+573105.9 & -1.0 & -0.6 & 1.2 & 18.28$\pm$0.15 & 1 & 0.177 & 0.133 & -23.93 \\
\multicolumn{9}{l}{Control Quasars}\\
J093411.14+000519.7 & -0.6 & -0.5 & 0.8 & 17.82$\pm$0.11 & 1 & ---  & ---   & ---    \\
J114445.93+055053.8 & 12.6 & 0.0 & 12.6 & 19.76$\pm$0.15 & 1 & ---  & ---   & ---     \\
J142519.70+012043.2 & 1.4  & 9.7 & 9.8  & 18.77$\pm$0.17 & 1 & ---  & ---   & ---     \\
J150306.34+011345.8 & 0.5  & 7.3 & 7.3  & 18.76$\pm$0.18 & 1 & ---  & ---   & ---     \\
J160108.43+422937.5 & -8.6 & 2.7 & 9.0 & 18.92$\pm$0.19 & 1 & ---  & ---   & ---     \\
\hline
\end{tabular}
\vspace*{-0.4cm}
\end{minipage}
\end{center}

\end{table*}      

\section{The $K$-band images of the absorber sight-lines}\label{sec:results}

\subsection{$K$-band number counts}\label{sec:nummag}

The total sky coverage of the 36 fields, each covering a $47\farcs8
\times 47\farcs8$ region, excluding a 1\farcs0 strip at the frame
edges where the image catalogues will be incomplete, is 22.8\,arcmin$^2$.
Excluding a 6\farcs0 radius circle centred on each of the
35 quasars, which will be used in subsequent analysis of galaxies
associated with the quasar absorber, reduces the area to 21.7\,arcmin$^2$.
The number of galaxies is relatively small but the
number-counts (Table \ref{tab:nm}) are consistent with the results of
the K20 survey and the compendium of $K$-band imaging results
presented by \citet{2002A&A...392..395C}. The contribution of stars to
the counts fainter than $K=18.5$ at the high Galactic latitudes of the
fields is expected to be small ($\la 5\,$per cent)
\citep{2002A&A...392..395C} and all objects fainter than $K=18.5$ are
treated as `galaxies' in subsequent statistical analysis. The large
number of widely-separated fields reduces the impact of large scale
structure on the number-counts but, nonetheless, the Poisson errors
associated with the observed numbers of galaxies underestimate the
uncertainties given the clustering of galaxies.

\begin{table}
\begin{center}
\caption{\label{tab:nm} Integral number-magnitude counts for the `field'
compared to the predictions from the K20 survey 
\citep{2002A&A...392..395C}.}
\begin{tabular}{ccc} \hline
Magnitude & $N_{obs}$ & $N_{K20}$ \\ \hline
$K \le 18.0$ & 28$\pm5$ & 44$\pm5$ \\
$K \le 19.0$ & 96$\pm10$ & 122$\pm8$ \\
$K \le 20.0$ & 221$\pm15$ & 204$\pm9$ \\
\hline
\end{tabular}
\vspace{-0.4cm}
\end{center}
\end{table}

\subsection{Galaxies at the quasar redshifts}

The selection of the absorber sample deliberately included a bias
towards quasars with redshifts well above those of the absorbers. A
small sample of quasars without known intervening absorption line
systems was also imaged to provide an empirical control sample.  The
median redshift of the quasar sample is $z_{median}$=1.65 and only
five quasars possess redshifts $z < 1.4$.

There are three independent results which consistently indicate a low
level of contamination of the number--magnitude counts by galaxies
physically associated with the quasars themselves.  Firstly, the
spectroscopic results from the K20 survey show only 9\,per cent of
their $K\le 20.0$ sample to possess redshifts $z > 1.5$
\cite{2002A&A...391L...1C}.  Secondly, the small control sample of
five quasars with $z \simeq 1.5$ contains just a single galaxy
detected within a radius of 6\farcs0.  Finally, the number of galaxies
detected within 6\farcs0 of the absorber sample quasars as a function
of quasar redshift also provides no evidence for significant
contamination: 16 galaxies were detected around the 15 quasars
with $z < z_{median}$, with 14 galaxies detected around the 15 quasars
with $z > z_{median}$; and of the five quasars with $z < 1.4$, two
have no detected galaxies.

\subsection{Quasar host galaxies}

The $K$-band magnitudes of host galaxies of quasars with $1 \le z \le
2$, which are not strong radio sources and possess absolute magnitudes
within the range covered by our sample, are typically $K \sim 19$
\citep[e.g.][]{2004ApJ...604..495F, 2006NewAR..50..772K}. While no
account of the radio properties of the quasars was taken when
selecting the sample, only one of the quasars (SDSS
J120300.96+063440.8 at $z$=2.182, with a ratio of radio to
$i$-band optical flux $R_i \simeq 3.3$) is a strong radio source
\citep[$R_i > 1.0$][]{2002AJ....124.2364I}.

Although the expected magnitudes of the quasar host galaxies lie above
the magnitude limit of our object catalogues ($K=20.0$), the size and
surface brightness properties of the host galaxies effectively
preclude their detection via the techniques employed to identify
potential absorber host galaxies.  The quasar host galaxies are
physically large, with half-light radii typically exceeding
$1\farcs0$, and the surface brightnesses are low, not least because of
the rapidly increasing impact of the $(1+z)^4$ cosmological dimming as
$z_{quasar}$ exceeds unity.  The $K$-band observations of
\citet{2004ApJ...604..495F} and others involve exposure times
equivalent to a factor $\sim$4 larger than those presented here and peak
surface brightnesses of the detected host galaxies are typically found
to be $\mu_K \ga 20\,$mag\,arcsec$^{-2}$ for quasars that are not
strong radio sources.

The almost complete lack of extended, approximately symmetric sources
surrounding the quasar images is thus not surprising given the much
higher surface brightness thresholds, $\mu_K \simeq
21.1\,$mag\,arcsec$^{-2}$, to which our imaging reaches.  It is
certainly possible that a number of small, faint, $K > 20.0$,
residuals centred on the PSF-subtracted quasar images could be the
higher surface brightness cores of host galaxies but the presence of
any such hosts does not result in contamination of the catalogue of
$K \le 20.0$ objects that forms the basis for the statistical
analysis of the absorber hosts.

\subsection{The companion galaxy catalogue}\label{sec:galcat}

\begin{figure*}
  \includegraphics[scale=0.95]{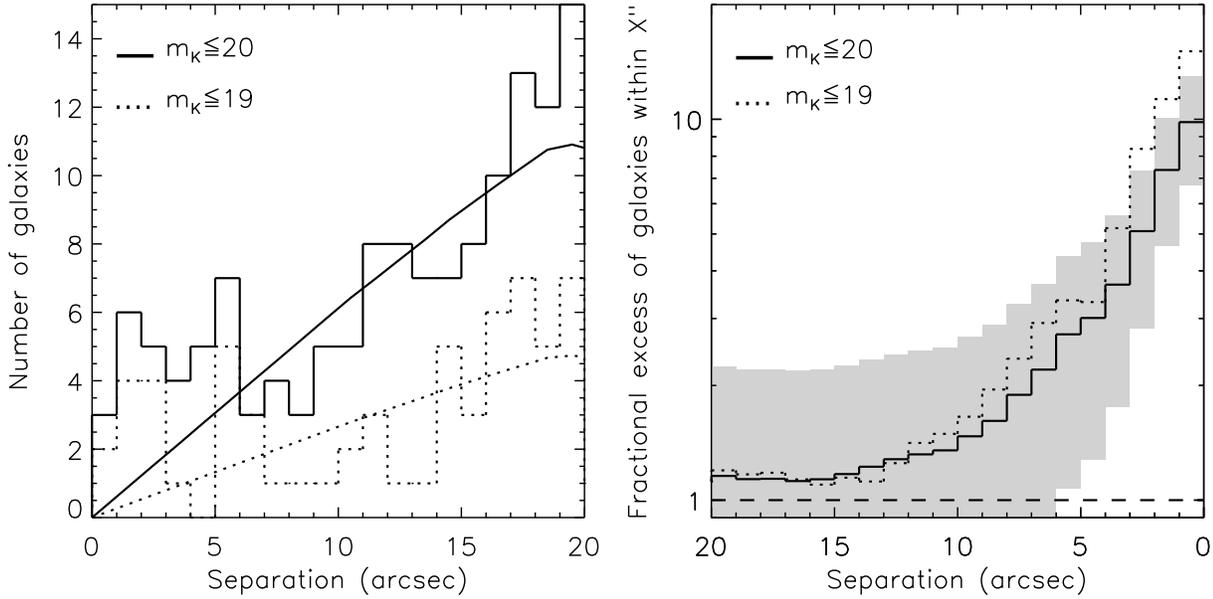}
  \caption{{\it Left:} Number of galaxies with $K\le20$ as a
  function of projected distance from the quasar, in bins of
  1\farcs0. {\it Right:} Cumulative number of galaxies with $K\le20$
  (full line), $K\le19$ (dotted line). In both figures, the
  histograms are the observations and the continuous lines are the
  predictions assuming no excess of galaxies around the position of
  the absorber.}\label{fig:galex}
\end{figure*}

The discussion in the preceding sub-sections indicate that the census
of galaxies with $K \le 20.0$ is essentially complete to within
1\farcs0 of the quasar images.  Fainter than $K \simeq 19.0$, the
census is substantially incomplete for galaxies with very small
$\la0\farcs5$ separations from the quasar.  The corresponding spatial
separations are small however, corresponding to scales
$\la$4\,kpc. The galaxy catalogue is also expected to suffer from
little contamination by quasar host galaxies or other galaxies at the
quasar redshifts, based both on prior expectations from larger
$K$-band surveys and on the lack of any empirical evidence from the
imaging observations themselves. The statistics of the sample of
galaxies presented in Table \ref{tab:galcat} should thus provide
direct constraints on any population of galaxies associated with the
\caii absorbers that are bright enough to be included in the $K \le
20.0$ sample.

The effectiveness of an imaging survey for constraining the presence
of associated galaxies depends on: i) the number of targets, ii) the
surface density of unrelated galaxies and iii) the number of
associated galaxies per target.  Figure \ref{fig:galex} presents the
number counts of galaxies detected in our survey, as a function of
separation from the absorber. On the right we show the observed number
of galaxies within a given radius, divided by that predicted; the grey
shading shows the associated Poissonian errors for the magnitude limit
of 20. For our observations, the angular scale on which such an excess
population could be identified with any significance is
$\simeq$5-7$^{\prime\prime}$ and we chose a radius of 6\farcs0 for the
presentation of the associated galaxy statistics. We have not
attempted to maximise the significance of the excess by optimising the
choice of magnitude limit and separation (e.g.  quoting statistics for
the excess at $<$3\farcs0 with a magnitude limit of $K\le19.0$).

The existence of a galaxy overdensity close to the quasars with
intervening \caii absorbers is clear. Table \ref{tab:galden}
summarises the number-counts within 6\farcs0 for the absorber and
control quasars down to magnitude limits of $K=19.0$ and $20.0$. The
predicted numbers of galaxies are calculated from the number counts of
galaxies at $\Delta\theta > 6\farcs0$ from the target quasar in all
target fields (Section \ref{sec:nummag}). The numerical excess of 
galaxies observed is high, with observed:predicted numbers of 30:9.6 and
16:4.2 for magnitude limits of $K=20.0$ and $19.0$
respectively. Figure \ref{fig:galex} illustrates the form of the
observed excess as a function of separation.  The quasars are widely
separated on the sky and large-scale structure is not an issue but the
small-scale clustering of galaxies may be expected to enhance the
apparent overdensity. However, allowing only {\it one} galaxy within a
6\farcs0 radius of each absorber quasar to contribute to our number
count of observed galaxies, gives observed:predicted numbers of 21:9.6
and 12:4.2 for magnitude limits of $K=20.0$ and $19.0$ respectively.
The significance of the excess at both magnitudes is $\simeq$3$\sigma$.
The statistics for the small control sample of quasars are entirely
consistent with the predictions from the field $K$-band number counts.


\section{Properties of the Absorber-Associated Galaxies}\label{sec:props}

\subsection{The galaxy luminosity function}

\begin{figure*}
  \includegraphics[scale=0.95]{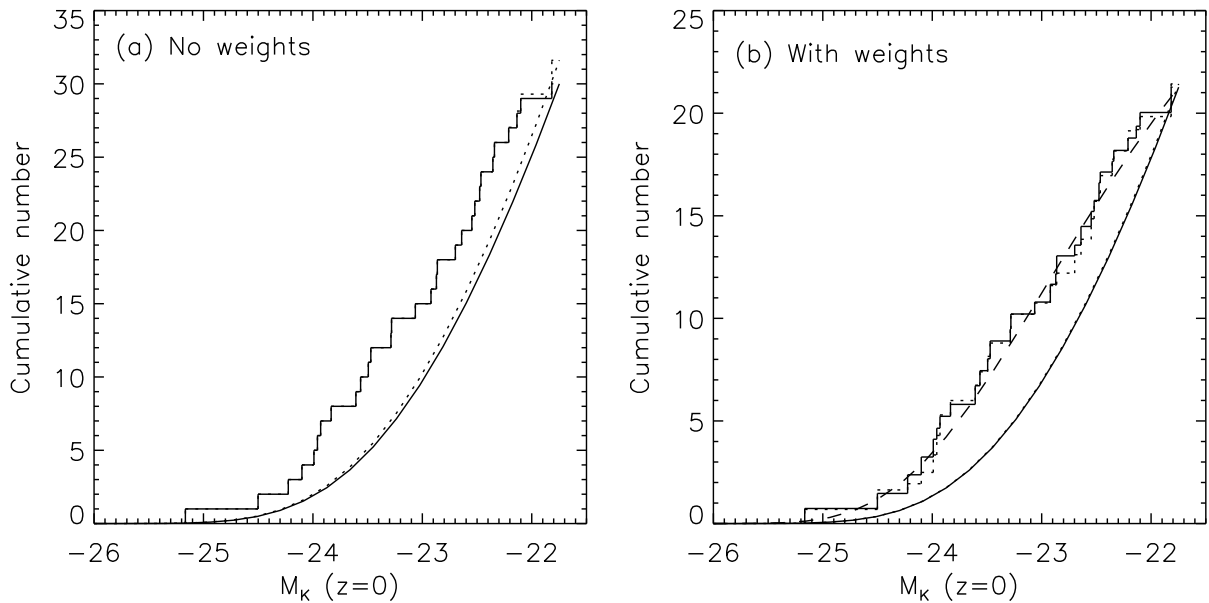}
   \caption{The $z=0$ cumulative galaxy luminosity function (GLF) of
   galaxies within 6\farcs0 from the absorber, assuming they lie at
   the redshift of the absorber (stepped histograms). {\it Left:} All
   galaxies are included, as observed. The dotted histogram accounts
   for the magnitude limit of the survey. {\it Right:} Two weighting
   schemes are applied to allow for interlopers, based on each
   galaxy's distance from the absorber (full histogram) or each
   galaxy's apparent magnitude (dotted histogram). The smooth
   overplotted lines are the $z=0$ $K$-band luminosity function of
   \citet{2001ApJ...560..566K}. In the right hand panel, the dashed
   line results from scaling this luminosity function by $L^{0.7}$,
   our best match to the \caii--selected galaxies (see text).}\label{fig:glf}
\end{figure*}

The form of the local $K$-band galaxy luminosity function (GLF) is now
well-established \citep[Table 4 of][]{2001ApJ...560..566K} and we
adopt the Schechter fit presented in this paper, with $M_K^*=-23.39$
(for $H_0=100$\,km\,s$^{-1}$) and $\alpha=-1.09$, as our fiducial
reference when considering the absolute magnitude distribution of the
galaxies apparently associated with the \caii absorbers.

The $K$-band observations probe the rest-frame wavelength region
$\sim$10\,000\,\AA \ ($z=0.7$) to $\sim$13\,000\,\AA \ ($z=1.2$) where
the galaxy light is dominated by K-giant stars.  As a result, the
K-corrections in the $K$-band are largely insensitive to the spectral
type of the galaxies.  Galaxies observed at increasing redshift are
expected to brighten, as their stellar populations become younger, and
there is broad agreement between the predictions of theoretical models
\citep[e.g.][]{1996MNRAS.281..953P} and observations
\citep{2003ApJ...595..698D, 2006MNRAS.367..349S}.  To characterise the
evolution we adopt a simple pure luminosity evolution model for the
GLF, with the characteristic absolute magnitude evolving as
$M_K^*(z)=M_K^*(z=0)-0.53z$, as determined by
\citet{2003ApJ...595..698D}.  Our observations probe a magnitude range
typically within $\pm1.5\,$mag of $M_K^*$ and possible variations in
the slope of the faint end slope of the GLF do not significantly
affect our results.  Cols.  2 and 3 of Table \ref{tab:galkcorr}
presents K-corrections as a function of redshift for galaxies of two
spectral types.  Absolute magnitudes before and after accounting for
redshift evolution are given for objects with $K=18.0$ and $20.0$ in
Cols.  4, 5, 7 and 8.  Finally, luminosities relative to $L^*$ are
provided in Cols.  6 and 9.  The $K$-band at $z$$\simeq$1 approximates
the $J$-band at $z$=0.0 and the K-corrections and luminosity evolution
are well-defined at the relatively long wavelengths probed by the
$K$-band observations.  Beginning with the $J$-band luminosity
function of \citet{2001MNRAS.326..255C} and using the observed
$K$-band GLF at $z$$\simeq$1 from \citet{2003A&A...402..837P} results
in luminosities for the galaxies, relative to $L^*$, that are
essentially indistinguishable from those calculated above.

A GLF for the galaxies nominally associated with the absorbers can be
constructed by calculating the absolute magnitude of each of the
galaxies where an excess is observed, assuming they lie at the
redshift of the absorber.  After correcting for the effects of
redshift evolution as described above, a $z$=0.0 GLF is constructed
which may be compared with the local GLF derived by
\citet{2001ApJ...560..566K}. There are two statistical corrections
that must first be made before proceeding.

Firstly, although the limiting absolute magnitude, corresponding to the
$K=20.0$ apparent magnitude limit, does not vary enormously over the absorber
redshift range, a number of the fainter galaxies detected in the lower redshift
absorber fields could not have been detected at the redshifts of all the
observed fields.  Each galaxy is therefore weighted by the factor $N_{\rm
Detect}/N_{\rm Field}$, where $N_{\rm Detect}$ is the number of absorber fields
in which a galaxy of given absolute magnitude is detectable and $N_{\rm
Field}=30$ is the total number of fields we observe\footnote{This correction is
directly equivalent to the common V$_{\rm max}$ correction employed in
calculating mean properties of galaxies in redshift surveys.}.

Secondly, from the field number density (Section \ref{sec:nummag}) we
expect 9.6 galaxies of the 30 galaxies detected within a 6\farcs0
radius of the background quasar not to be associated with the
absorber, and we would like to account for these as best we can. In
what follows we designate $P_{\rm Field}$ as the probability that the
galaxy is {\it not} associated with the absorber; the average value is
$<P_{\rm Field}>=9.6/30=0.32$. We can use our prior knowledge of the
number-density and number-magnitude counts of the field galaxies to
predict for each of our galaxies a more accurate probability that it
is an interloping field galaxy, and then weight the contribution of
each galaxy to the GLF by individual $1-P_{\rm Field}$ factors. While
we could combine the priors into a single weighting scheme, we chose
to adopt algorithms based on each prior separately, to ensure that the
GLF is not sensitive to the precise scheme for the assignment of the
probabilities. Combining the weighting schemes gave a consistent
result.

The first weighting scheme accounts for the fact that the excess of
galaxies around the target quasar is observed to be a function of
distance from the quasar.  The individual $1-P_{\rm Field}$ factors
then take account of the increasing likelihood that a galaxy closer to
the quasar is associated with an absorber than one further away.
Galaxies in each annulus
($\Delta\theta$=0\farcs0--1\farcs0,1\farcs0--2\farcs0,...,5\farcs0- 
-6\farcs0)
are assigned a value of:
\begin{equation}
P_{\rm Field}=n_{\rm Field}/n_{Obs}
  \label{eq:psep}
\end{equation}
where $n_{\rm Field}$ and $n_{Obs}$ are the predicted and observed
number of galaxies within the annulus respectively.  The second scheme
is based on our prior knowledge of the apparent magnitude distribution
of $K$-band selected galaxies. Adopting the form
of the $K$-band number counts from \citet{2002A&A...392..395C},
magnitude limits of $K$=15.0 and 20.0, and ordering the galaxies,
$i$=1,2,...,30, by increasing magnitude, $K$, galaxies are assigned
a value of:
\begin{eqnarray}
P_{\rm Field, i} \propto n_{\rm Field}(15\le m \le m_i+0.5(m_{i+1}-m_i))  
\nonumber \\
-n_{\rm Field}(15 \le m < m_i-0.5(m_i-m_{i-1}))
\label{eq:pmag}
\end{eqnarray}
where $n_{\rm Field, i}$ is the predicted number of galaxies with
$K$-band apparent magnitudes between the limits for galaxy $i$ and the
results are normalised such that $<P_{\rm Field}>=0.32$.

Figure \ref{fig:glf}$a$ shows the $z$=0 cumulative GLF obtained without
applying the weighting schemes: the continuous histogram uses the raw
number counts; weighting galaxies by $N_{\rm Detect}/N_{\rm Field}$
results in the dotted histogram i.e. a very small correction for the
faintest objects.  The overplotted lines indicate the cumulative GLF
obtained by \citet{2001ApJ...560..566K}, normalised appropriately for
each histogram. Figure \ref{fig:glf}$b$ shows the cumulative GLF
obtained after applying the two different weighting schemes: the
continuous histogram uses weights according to spatial distribution
and the dotted histogram according to apparent magnitude. Clearly the
shape of the GLF is independent of the weighting scheme adopted for
the galaxies.  Again, the predicted GLF is overplotted, normalised
appropriately for each histogram.  The dashed line will be discussed
in Section \ref{sec:ldcc}. The lack of any particularly
large vertical displacements in any of the histograms shows that the
form of the cumulative GLF is not dominated by a small number of
galaxies with large weights.

There is a clear excess in the number of galaxies brighter than
$M_K^*$, reaching a factor $\simeq2$ at $M_K=-23.5$ and increasing to
a factor of $\simeq3$ at $M_K=-24.0$.

\begin{table*}
    \caption{\label{tab:galkcorr} K-corrections and absolute magnitudes 
    ($H_0=100$\,km\,s$^{-1}$) for $K_{\rm MKO}$-band
    observations of \citet{1980ApJS...43..393C} E- and Scd-template galaxy spectra as extended
    by \citet{2000A&A...363..476B}. The absolute magnitudes are calculated using the average
    of the E and Scd K-corrections at each redshift.
    }
\begin{tabular}{ccccccccc} \hline
z    & E K-correction & Scd K-correction & \multicolumn{3}{c}{$K=18.0$} & \multicolumn{3}{c}{$K=20.0$}  \\ 
     &          &         & $M_K(z=0)$ & $M_K(z)$ & $L_K(z)/L_K^*(z)$ & $M_K(z=0)$ & $M_K(z)$ & $L_K(z)/L_K^*(z)$ \\
\hline
 0.60 &  -0.598 &  -0.680 &  -23.32 &  -23.64 &   0.93 &  -21.32 &  -21.64 &   0.15 \\ 
 0.70 &  -0.641 &  -0.727 &  -23.69 &  -24.06 &   1.31 &  -21.69 &  -22.06 &   0.21 \\ 
 0.80 &  -0.678 &  -0.771 &  -24.00 &  -24.42 &   1.74 &  -22.00 &  -22.42 &   0.28 \\ 
 0.90 &  -0.707 &  -0.810 &  -24.28 &  -24.76 &   2.25 &  -22.28 &  -22.76 &   0.36 \\ 
 1.00 &  -0.739 &  -0.851 &  -24.53 &  -25.06 &   2.83 &  -22.53 &  -23.06 &   0.45 \\ 
 1.10 &  -0.757 &  -0.884 &  -24.76 &  -25.34 &   3.50 &  -22.76 &  -23.34 &   0.55 \\ 
 1.20 &  -0.752 &  -0.908 &  -24.99 &  -25.63 &   4.33 &  -22.99 &  -23.63 &   0.69 \\ 
 1.30 &  -0.736 &  -0.923 &  -25.20 &  -25.89 &   5.25 &  -23.20 &  -23.89 &   0.83 \\ 
 1.40 &  -0.699 &  -0.932 &  -25.41 &  -26.15 &   6.37 &  -23.41 &  -24.15 &   1.01 \\ 
 1.50 &  -0.669 &  -0.944 &  -25.61 &  -26.41 &   7.66 &  -23.61 &  -24.41 &   1.21 \\ 
 1.60 &  -0.635 &  -0.954 &  -25.79 &  -26.64 &   9.04 &  -23.79 &  -24.64 &   1.43 \\ 
 1.70 &  -0.603 &  -0.962 &  -25.97 &  -26.87 &  10.67 &  -23.97 &  -24.87 &   1.69 \\ 
 1.80 &  -0.584 &  -0.967 &  -26.13 &  -27.08 &  12.36 &  -24.13 &  -25.08 &   1.96 \\ 
 1.90 &  -0.564 &  -0.971 &  -26.28 &  -27.29 &  14.19 &  -24.28 &  -25.29 &   2.25 \\ 
 2.00 &  -0.550 &  -0.974 &  -26.42 &  -27.48 &  16.14 &  -24.42 &  -25.48 &   2.56 \\ 
\hline
\end{tabular}
\vspace*{-0.4cm}
\end{table*}     

\subsection{Galaxy morphologies}

Given the signal-to-noise ratio of the target imaging was set in order
to achieve detections of $K\simeq 20$ galaxies the quality of the
images is not in general sufficient to provide much quantitative
information on the morphologies of the galaxies. Visual inspection of
Fig. 1 and the {\sc SExtractor}-ellipticities indicate the presence of
a proportion of elongated images, suggesting that at least some of the
galaxies possess luminous disk components. There are certainly
examples of high surface brightness galaxies with substantial bulge
components (e.g. SDSS J132323.78-002155.2, SDSS J173559.95+573105.9)
and overall the $K$-band images provide no strong evidence for a
distribution of morphological properties that differ from those
evident for the general population of $\sim$$M^*$ galaxies at
$z$$\sim$1 in imaging surveys.

A link between \caii absorbers and interacting systems has been suggested
\citep{1991MNRAS.251..649B}.  Without redshifts to establish which of the galaxies
among the seven absorbers with more than one galaxy within 6\farcs0 are associated
with the absorbers, it is difficult to perform a strong test for an excess of close
pairs.

\subsection{The projected separation of the absorbers and the associated 
galaxies}

The angular separations of the centers of the galaxies from the
absorbers are given in Col.  4 of Table \ref{tab:galcat}.  By assuming
each galaxy is at the associated absorber redshift and converting the
angular separations into kpc, we can calculate the typical impact
parameters.  Again, the galaxies are weighted using the two different
weighting schemes discussed above.  The median projected distance
using separation- and magnitude-based weighting is 22.7\,kpc and
25.3\,kpc respectively, with the value for the weighting based on
separation smaller as expected. However, the relatively small
systematic difference gives confidence in the results and we adopt a
value of 24$\pm$2\,kpc for the median projected separation between the
\caii absorbers and the associated galaxies.

With chance projections constituting a third of the galaxy images within 
6\farcs0 of the absorbers extracting quantitative information concerning the
relationship between galaxy luminosity, impact parameter and absorber 
properties is not really viable. Inspection of the distribution of such
parameters do not reveal any convincing trends but definitive conclusions 
must await the acquisition of redshifts for the galaxies.

\subsection{The missing galaxies}

The statistical excess of galaxies about the \caii absorber quasars establishes
an association between the \caii absorbers and the observed galaxies but for
approximately a third of the absorbers no associated galaxy is seen.  Potential
explanations for the nature of the `missing' galaxies fall into two categories.
In the first, a residual concern for all imaging surveys of quasar absorber
fields, {\it all} the galaxies directly responsible for the absorbers are in
fact faint dwarf or low surface brightness galaxies, and the bright galaxies we
observe are simply neighbours associated through clustering.  In the second
category, the missing galaxies form part of the same galaxy population as those
producing the statistical excess, but limitations imposed by the sensitivity of
the observations result in the non--detections of the fainter or more--distant
galaxies.  We look at each option in turn.

\subsubsection{All absorber galaxies are undetected dwarf galaxies}

Could the galaxies directly associated with the absorbers {\it all} be
faint ($M_K \ga M_K^*+1.5$), likely possess small impact parameters,
and thus remain undetected for all the absorbers?  In such a model the
observed excess of luminous galaxies within 6\farcs0 arises from the
strong spatial clustering of galaxies on $\la$50\,kpc scales.  An
obvious example to illustrate this hypothesis is the Local Group, in
which absorber sightlines passing through the Magellanic clouds would
be associated from afar with the Milky Way or Andromeda. Indeed
similar scenarios have often been invoked to explain the low metallicities and
dust contents of DLAs.

Observationally, the determination of the clustering amplitude of dwarf galaxies
close to $\sim$$M^*$-galaxies is difficult and the constraints remain uncertain
\citep[e.g.][]{1980ApJ...238L..13L, 2005MNRAS.356.1233V, 2006ApJ...647...86C}.  At
$z\sim1$, local results may not apply in anycase.  Theoretically, Cold Dark Matter
simulations continue to overpredict the number of satellite galaxies
\citep{2004MNRAS.352..535D}, but additional processes may be important in
determining the frequency of satellites \citep[e.g.][]{2006MNRAS.368..563M}.

We will return to this scenario in the discussion of the true nature
of the \caii absorbers, suggesting the type of observations required to
definitively confirm it or otherwise.

\subsubsection{Only a fraction of absorber galaxies remain undetected}
\label{sec:ldcc}

There are three options for hiding the remaining third of the galaxies
that are unobserved: (i) they lie at more than 6\farcs0 from the
absorber; (ii) they are hidden by the PSF of the background quasar; or
(iii) they are fainter than the survey limit. Here we discuss each
scheme, to determine whether any option allows the unobserved galaxies
to form part of the same galaxy population that we do observe.

{\bf (i) They are distant from the absorber.} The finite size of our
sample means that sensitivity to the presence of excess galaxies at
separations beyond $\sim$50\,kpc is limited.  However, the statistics
of the galaxy surface density in the annulus 6\farcs0-12\farcs0
($\simeq$50-100\,kpc) surrounding the \caii absorbers show no
evidence for any excess (Table \ref{tab:galden}, Figure
\ref{fig:galex}), with observed:predicted numbers of 11:12.5
($K\le19.0$) and 28:28.8 ($K\le20.0$). Although a small number of
galaxies may indeed fall outside of the 6\farcs0 annulus, it is not
possible to explain the majority of the missing galaxies in this way.

{\bf (ii) They are hidden by the quasar PSF.} As discussed in Section
\ref{sec:psfsub}, the detectability of even relatively bright,
$K$$\simeq$19.0, galaxies declines rapidly at very small
separations, $\Delta\theta \la$0\farcs25 and the possibility that a
substantial fraction of the absorbers occur within impact parameters
of $\la$3\,kpc of the centers of intermediate luminosity galaxies
cannot be ruled out directly. However, the distribution of projected
absorber-galaxy separations for $\Delta\theta \ge 0\farcs5$ shows no
evidence for a rapid increase in the frequency of galaxies at very
small impact parameters. A single population drawn from the same
$(\Delta\theta$,$M_K)$ distribution as observed is thus ruled out,
although a discontinuity in the distribution of absorber--galaxy
separations is allowed by the observations.

{\bf (iii) They are faint.}  Perhaps the missing galaxies possess a similar
distribution of absorber-galaxy separations but have magnitudes fainter than the
$\simeq$$M_K^*+1.5$ limit of the $K$-band images?  Previous studies of the
absorption cross--sections of galaxies \citep[e.g.][]{1994ApJ...437L..75S}, have
found a luminosity dependence of the observed size of absorbing regions
associated with galaxies of the form,

\begin{equation}
\sigma(L)=\sigma^* \left( \frac{L}{L^*} \right) ^{\beta}
  \label{eq:absrad}
\end{equation}
where $\sigma^*$ is the cross-section for an $L^*$ galaxy. Adopting such a
luminosity dependent scaling for the \caii absorbers, with $\beta=0.7$,
produces an excellent fit to the galaxy luminosity function for the galaxies
brighter than the $\simeq$$M^* +1.5$ limit of our observations (Fig. 
\ref{fig:glf}). Given the shallow slope, $\alpha=-1.09$, of the 
\citet{2001ApJ...560..566K} GLF, combined with the strong bias towards luminous
galaxies, the integrated cross--section of the galaxy population converges
rapidly at faint magnitudes. Adopting a faint limit of $M^* + 5$ for the
integration results in a prediction that $\simeq$60\, per cent of the
absorber cross--section should result from galaxies brighter than our
observational limit of $M^* +1.5$, in good agreement with the fraction of
our 30 absorbers where we see such a galaxy. Given the power law 
luminosity--dependent scaling provides an effective and simple 
parameterisation of our results, we adopt the relation in order to make
quantitative estimates of the filling-factor of the \caii absorption
and to compare the properties of the \caii absorbers with other classes
of quasar absorption systems.

\begin{table}
\begin{center}
\caption{\label{tab:galden} Companion galaxy surface number densities
  in the 30 absorber and five control fields, as a function of separation
  from the quasar. Counts for galaxies with $K\le 20$ and $K\le
  19$ are both given.}
\begin{tabular}{ccccc} \hline
Separation & Observed & Predicted & Observed & Predicted \\
  (")     &
\multicolumn{2}{c}{Absorbers} &
\multicolumn{2}{c}{Controls} \\
\hline
\multicolumn{2}{l}{(a) \ \ \ $K \le 20.0$} \\
$\le$1.0 & 3  & 0.3 & 1 & $<$0.1 \\
$\le$2.0 & 9  & 1.1 & 1 &  0.2 \\
$\le$3.0 & 14 & 2.4 & 1 &  0.4 \\
$\le$4.0 & 18 & 4.3 & 1 &  0.7 \\
$\le$5.0 & 23 & 6.7 & 1 &  1.1 \\
$\le$6.0 & 30 & 9.6 & 1 &  1.6 \\
 & & & & \\
6.0--12.0 & 28 & 28.8 & 5 & 4.8 \\
 & & & & \\
\multicolumn{2}{l}{(b) \ \ \ $K \le 19.0$} \\
$\le$1.0 & 2  & 0.1 & 1 & $<$0.1 \\
$\le$2.0 & 6  & 0.5 & 1 &  0.1 \\
$\le$3.0 & 10 & 1.0 & 1 &  0.2 \\
$\le$4.0 & 11 & 1.9 & 1 &  0.3 \\
$\le$5.0 & 11 & 2.9 & 1 &  0.5 \\
$\le$6.0 & 16 & 4.2 & 1 &  0.7 \\
 & & & & \\
6.0--12.0 & 11 & 12.5 & 2 & 2.1 \\
\hline
\end{tabular}
\vspace*{-0.4cm}
\end{center}
\end{table}

 \begin{table*}
    \caption{\label{tab:caiiprop} Absorber properties: rest-frame equivalent
    widths ($W$) of \caii, \mgii, \mgi and \feii absorption lines and
    errors. The final column gives neutral Hydrogen column densities
    from \citet{2006ApJ...636..610R}. }
\begin{minipage}{18.5cm}
\begin{tabular}{ccccccccccc} \hline
Name & $z_{abs}$ & $W_{\lambda3935,3970}$ & err &
$W_{\lambda2796,2803}$ & err & $W_{\lambda2853}$ & err  & $W_{\lambda2600}$ & err & 
N(HI) \\
\hline
J074804.06+434138.5 & 0.898 & 0.62,0.53 & 0.23,0.16 & 1.69,1.17 & 0.12,0.08 & 0.30 & 0.13 & 0.68 & 0.16 & --- \\
J080735.97+304743.8 & 0.969 & 0.65,0.77 & 0.16,0.14 & 2.51,2.40 & 0.13,0.09 & 1.21 & 0.14 & 2.15 & 0.15 & --- \\
J081930.35+480825.8 & 0.903 & 0.68,0.34 & 0.12,0.07 & 1.66,1.53 & 0.08,0.05 & 1.00 & 0.10 & 1.39 & 0.10 & --- \\
J083819.85+073915.1 & 1.133 & 0.71,1.15 & 0.26,0.41 & 2.63,2.42 & 0.13,0.09 & 0.67 & 0.16 & 1.70 & 0.15 & --- \\
J085221.25+563957.6 & 0.844 & 0.66,0.49 & 0.22,0.18 & 3.29,3.02 & 0.16,0.11 & 1.18 & 0.23 & 2.52 & 0.18 & --- \\
J085556.63+383232.7 & 0.852 & 0.46,0.07 & 0.13,0.07 & 2.68,2.53 & 0.09,0.06 & 0.70 & 0.13 & 2.30 & 0.10 & --- \\
J093738.04+562838.8 & 0.978 & 1.23,0.53 & 0.28,0.19 & 4.88,4.27 & 0.25,0.20 & 2.42 & 0.31 & 3.51 & 0.30 & --- \\
J095307.05+111140.8 & 0.980 & 0.88,0.69 & 0.30,0.25 & 2.34,2.33 & 0.21,0.15 & 1.84 & 0.33 & 2.05 & 0.32 & --- \\
J095352.69+080103.6 & 1.023 & 0.48,0.35 & 0.13,0.09 & 0.85,0.76 & 0.07,0.05 & 0.40 & 0.21 & 0.58 & 0.08 & --- \\
J100339.44+101936.8 & 0.816 & 0.76,0.57 & 0.23,0.15 & 1.46,1.31 & 0.15,0.11 & 0.55 & 0.19 & 1.35 & 0.17 & --- \\
J105930.50+120532.8 & 1.050 & 1.85,0.66 & 0.39,0.17 & 1.71,1.97 & 0.24,0.18 & 0.94 & 0.39 & 2.11 & 0.26 & --- \\
J110729.03+004811.2 & 0.740 & 0.41,0.22 & 0.06,0.04 & 2.89,2.79 & 0.04,0.02 & 0.85 & 0.05 & 2.33 & 0.04 & 21.00$^{+0.02}_{-0.05}$ \\
J113357.55+510844.9 & 1.029 & 1.16,0.58 & 0.40,0.24 & 2.65,2.73 & 0.21,0.17 & 0.75 & 0.29 & 1.94 & 0.23 & --- \\
J115523.97+480141.6 & 1.114 & 2.21,1.45 & 0.70,0.49 & 3.67,3.15 & 0.36,0.26 & 1.33 & 0.30 & 3.84 & 0.45 & --- \\
J120300.96+063440.8 & 0.862 & 1.34,1.05 & 0.28,0.19 & 5.65,5.04 & 0.35,0.27 & 3.03 & 0.49 & 3.77 & 0.53 & --- \\
J122144.62-001141.8 & 0.929 & 0.59,0.29 & 0.18,0.11 & 0.99,0.84 & 0.13,0.09 & 0.75 & 0.20 & 0.71 & 0.15 & --- \\
J122756.35+425632.4 & 1.045 & 0.34,0.17 & 0.10,0.06 & 1.58,1.37 & 0.07,0.05 & 0.25 & 0.12 & 1.14 & 0.07 & --- \\
J124659.81+030307.6 & 0.939 & 0.95,0.58 & 0.36,0.16 & 2.92,2.80 & 0.19,0.14 & 1.28 & 0.22 & 2.13 & 0.26 & --- \\
J130841.19+133130.5 & 0.951 & 0.73,0.70 & 0.26,0.20 & 1.82,1.63 & 0.18,0.13 & 0.73 & 0.20 & 1.09 & 0.22 & --- \\
J132323.78-002155.2 & 0.716 & 0.95,0.55 & 0.10,0.07 & 2.16,1.98 & 0.11,0.07 & 0.97 & 0.13 & 1.44 & 0.13 & 20.54$^{+0.15}_{-0.15}$ \\
J132346.05-001819.8 & 1.087 & 1.07,0.54 & 0.37,0.38 & 4.24,4.06 & 0.53,0.40 & 2.00 & 0.63 & 2.98 & 0.36 & --- \\
J140444.19+551636.9 & 1.070 & 0.94,0.50 & 0.36,0.22 & 1.96,1.83 & 0.26,0.19 & 0.68 & 0.25 & 1.42 & 0.23 & --- \\
J145633.08+544831.6 & 0.879 & 0.42,0.40 & 0.16,0.10 & 4.01,3.72 & 0.12,0.08 & 1.66 & 0.19 & 3.09 & 0.15 & --- \\
J151247.48+573843.4 & 1.044 & 1.00,0.53 & 0.26,0.16 & 2.02,2.27 & 0.20,0.14 & 0.73 & 0.38 & 1.59 & 0.23 & --- \\
J153503.36+311832.4 & 0.904 & 0.77,0.26 & 0.15,0.10 & 2.16,1.86 & 0.12,0.09 & 0.77 & 0.16 & 0.98 & 0.12 & --- \\
J155529.40+493154.9 & 0.893 & 0.38,0.18 & 0.12,0.07 & 2.25,2.06 & 0.08,0.06 & 0.71 & 0.09 & 1.53 & 0.11 & --- \\
J160932.95+462613.3 & 0.966 & 0.75,0.28 & 0.28,0.16 & 1.01,0.86 & 0.15,0.10 & 0.52 & 0.35 & 0.69 & 0.16 & --- \\
J164350.94+253208.8 & 0.885 & 0.54,0.13 & 0.20,0.13 & 0.85,0.65 & 0.09,0.06 & 0.19 & 0.08 & 0.45 & 0.09 & --- \\
J172739.01+530229.2 & 0.945 & 0.62,0.61 & 0.15,0.12 & 2.69,2.56 & 0.10,0.07 & 0.99 & 0.13 & 2.23 & 0.12 & 21.16$^{+0.04}_{-0.05}$ \\
J173559.95+573105.9 & 0.872 & 0.80,0.68 & 0.18,0.12 & 1.98,1.76 & 0.12,0.08 & 0.83 & 0.15 & 1.53 & 0.16 & --- \\
\hline
\end{tabular}
\vspace*{-0.4cm}
\end{minipage}
\end{table*}      

\section{Discussion}\label{sec:disc}

\subsection{The \caii absorbers and their associated galaxies}

The $K$-band imaging results of the \caii systems provide new
constraints on the relationship between \caii absorbers and their
associated galaxies. In \citet[][Section 4.1]{2007MNRAS.374..292W} we
used observations of DLAs, together with the relative number densities
of DLAs and \caii absorbers, to propose that \caii absorption can
arise at projected radii of $\sim10$\,kpc from the centre of a
galaxy. This simple calculation assumes a circular geometry for the
absorbers and a unit filling factor (i.e. the gas is not
patchy). Clearly, this model contrasts with the much larger value of
24\,kpc derived from the $K$-band imaging. Our imaging results reveal
that the dominant contributor to the \caii cross-section is {\it not} the
inner part of more extended DLA absorbers, centred on relatively
luminous galaxies.  Rather, the \caii absorbers themselves trace much
larger structures associated primarily with luminous galaxies where the 
filling factor is low.

Taking the value of $dP/dz=0.025$ from \citet{2007MNRAS.374..292W} for
\caii absorbers at $z$$\simeq$1 and integrating the
luminosity-dependent absorber cross--section (Section \ref{sec:ldcc})
using the \citet{2001ApJ...560..566K} GLF down to $M^*+5$ produces a
value of $\sigma^*=400$\,kpc$^2$, equivalent to a radius,
$R^*=11.4$\,kpc.  Our observations produce an observed mean
galaxy-absorber separation of 24\,kpc for galaxies with $M=M^*$,
leading to a maximum radius out to which an absorber may be
detected\footnote{The mean observed impact parameter is about
two-thirds of the maximum radius, assuming the absorbers present
simple circular cross-sections.}  of 36\,kpc.  The inferred
filling-factor is thus only (11.4/36)$^2$=0.1.

The large mean projected separations observed also provide an
explanation for the small contribution of the \caii absorbers to the
star formation rate density at $z\sim1$ determined by
\citet{2007MNRAS.374..292W} from \oii$\lambda\lambda$3727,3730
emission observed in the SDSS spectra. The fraction of the $K$-band
luminosity of the associated galaxies that would fall within the
3\farcs0 diameter of the SDSS spectroscopic fibres is only
30$\pm5$\,per cent. Thus, the star formation associated with the
central luminous parts of the associated galaxies would have been
missed and, assuming the observed galaxies are directly responsible
for the absorption, the results of \citet{2007MNRAS.374..292W} do not
constrain directly the contribution of \caii absorber galaxies to the star
formation rate of the Universe at $z\sim1$.

\subsection{Comparison with results for other absorber classes}

In this section we compare our imaging results to those for the more
familiar classes of strong \mgii absorbers and DLAs.

\subsubsection{\mgii\ absorbers}

The \mgii equivalent widths of the \caii absorbers are very large; the mean
rest-frame \mgii $W_{\lambda2796}$ of the sample imaged in this paper is
2.44\,\AA, with a minimum value of 0.85\,\AA, and 18 of the 30 systems have
equivalent widths greater than 2.0\,\AA \ (Table \ref{tab:caiiprop}).  Where
possible, we therefore focus on results for similarly strong \mgii absorbers.

The only published galaxy luminosity function for \mgii absorbers is
from the survey of 52 absorbers by \citet{1994ApJ...437L..75S}, who
find that, after correcting for an observed dependence of gas
cross-section on galaxy luminosity ($\sigma\propto L^{0.4}$), the
$K$-band luminosity function of \mgii absorbers in the redshift range
$0.2\le z \le1.0$ is consistent with the $z$=0 luminosity function of
\citet{1993MNRAS.263..560M}.  This implies a weaker galaxy
luminosity-dependence of the cross-section than for \caii
absorbers. Several caveats apply to the comparison with Steidel et
al.'s result.  Firstly, the \mgii equivalent width distribution of the
absorber sample differs considerable from ours, with their systems
selected to have $W_{\lambda2796} >$0.3\,\AA.  Secondly, a recent
reassessment of the identification procedure for host galaxies in this
survey by \citet{2005ASPC..331..387C}, has revealed a small number of
potential systematic biases which remain to be investigated.

Our second comparison is with the $r$-band imaging of the environments of 15
very strong, $W_{\lambda2796} \ge 2.7$\,\AA, \mgii\ absorbers with redshifts
$0.42 < z < 0.84$ by \citet{2006astro.ph.10760N}.  The principal result is the
detection, in essentially all cases, of a galaxy, which, if physically
associated with the absorber, lies projected within 40\,kpc of the quasar and
has luminosity $L \ga 0.3L^*$.  Seven of the sample are \caii absorbers, and
six of these possess a galaxy within 40\,kpc with luminosities $0.3L^*
\la L \la 2.0L^*$. Allowing for the chance projection of one or two galaxies,
the results are entirely consistent with the findings presented in this paper.

The second result of the Nestor et al.  study is the detection of an
excess of apparently bright galaxies, which, if physically associated
with the absorber, have projected separations extending to
$\sim$150\,kpc and extremely high luminosities, $4L^* \la L \la
13L^*$. We can perform a similar analysis using our dataset by comparing
the observed number of galaxies brighter than $4L_K^*(z)$ if placed at
$z_{abs}$ in each of our \caii absorber field images (which extend to
$\simeq$150\,kpc from the quasars) to the predictions from the
K20-survey number-magnitude distribution.  Eight field galaxies are
predicted, whereas we find a total of 11 such galaxies around
eight quasars.  Thus, we find no evidence in our sample for any such
excess.  Nestor et al.  discuss an interpretation for the observed
excess in terms of elevated levels of star-formation in the
galaxies. At the rest-frame wavelengths of $\sim$4000\,\AA \ probed by
their observations, they are certainly far more affected by the
presence of ongoing and recent star-formation than the
$\sim$10\,000\,\AA \ wavelengths probed by our $K$-band observations.
However, we note that the {\tt PhotoZ} photometric redshifts from the
SDSS DR5 (Adelman-McCarthy et al.  2007) for all of their galaxies
with $L \ge 4L^*$ are consistent with their lying at considerably lower
redshifts than the absorbers, supporting their identification as more
typical objects at lower redshifts\footnote{Such an interpretation
does not explain why such an excess of galaxies at redshifts $z_{gal}
<< z_{abs}$ is observed}.

Finally, a powerful constraint on the luminosities of galaxies associated with
\mgii absorbers comes from the stacking analysis of SDSS images in the regions
around \mgii absorbers by \citet{2006astro.ph..9760Z}.  The mean integrated
rest-frame luminosity within $100$\,kpc of their sample of \mgii absorbers with
redshifts $0.76 \le z \le 1.0$ is $M_i(z)=-22.4$ (AB system).  The absolute
magnitude corresponds to $\simeq$$M^*-0.5$, or $M^*$($z$=0).  At lower redshifts,
where the signal is strong enough for division of the sample,
\citet{2006astro.ph..9760Z} find only a small difference between the mean
absolute magnitudes of galaxies associated with all the \mgii absorbers and
those associated with the strongest absorbers.  Assuming the same holds for the
$0.76 \le z \le 1.0$ absorber sample, and noting that approximately half the
luminosity within 100\,kpc lies within the smaller aperture of 50\,kpc, produces
an average absolute magnitude of $\simeq$$M^*$($z$=0)+0.75.  The 50\,kpc aperture
corresponds closely to the spatial scale of our observed excess of galaxies and
the average absolute magnitude of galaxies associated with all 30 \caii
absorbers is $\simeq$$M^*$($z$=0)+0.25$\pm$0.2.  The estimate is insensitive to
whether the galaxies associated with the nine absorbers without detected
galaxies are excluded or included (even assuming detections at the $K$=20.0 limit).
The somewhat brighter mean galaxy luminosity associated with the \caii
absorbers is consistent with the stronger galaxy luminosity-dependence of the
cross-section ($\sigma \propto L^{0.7}$) compared to that for \mgii absorbers
($\sigma \propto L^{0.4}$).

\subsubsection{DLA absorbers}

Notwithstanding recent efforts \citep{2006ApJ...636..610R}, the number of
known DLA-systems at redshifts $z$$\la$1, where imaging studies of host
galaxies are most effective, is small.  \citet{2005ApJ...620..703C} provide a
summary of relevant imaging studies and four additional systems have been imaged
by \citet{2006AJ....131..686C}.  Follow-up studies of a small number of the
galaxies associated with the DLAs have also been undertaken \citep[e.g.  Chen et
al.  2005;][]{2007AJ....133..130G}.  Even from the limited statistics available
it is clear that the galaxies associated with DLAs exhibit an extended range in
the distributions of both luminosity and impact parameter.  Our results for
the three confirmed DLAs (Table \ref{tab:caiiprop}) reinforce the conclusion,
with one non-detection, one luminous galaxy with a small (8\,kpc) impact
parameter and one intermediate luminosity galaxy with a significant (30\,kpc)
impact parameter.  The statistics are as yet too poor to allow strong
conclusions to be drawn but the observed distribution of luminosity and impact
parameter for the galaxies associated with the \caii absorbers is consistent
with what is known for DLAs.

\section{Concluding remarks}\label{sec:concl}

As discussed in Section \ref{sec:intro}, the nature of quasar absorption line
systems is still a matter of considerable debate.  The Steidel et~al.  (1994)
model of \mgii absorbers with $W_{\lambda2796}>0.3$\,\AA \ residing within
spherical halos of radius $\simeq$40\,kpc and unity filling factor has strongly
influenced our picture of quasar absorption line systems for the past decade.
However, alternative models have been suggested \citep[e.g.][]{1996ApJ...465..631C} and
recent imaging results indicate that substantially larger impact parameters are
common and filling factors less than unity are necessary
\citep{2005pgqa.conf...24C, 2006astro.ph..9760Z}.

Theories range from identifying absorption systems with:  extended disks, large
gaseous haloes, outflows from superwinds, or, extended debris from
mergers/interactions.  Although models are currently not well advanced,
determining the relation between absorption line systems (with different metal
line and dust properties) and the properties of their hosts (impact parameters,
luminosity and morphologies), should lead to a better understanding of the
origins of absorption line systems and their relevance to galaxy evolution.

In Wild et~al.  (2006) we found the dust content of \caii absorption
line systems to differ from those of \mgii absorption line systems in
the same redshift range and with the same \feii and \mgii equivalent
width properties.  More specifically, the \caii absorption line
systems have \EBV\ = 0.06 compared to 0.008 for the entire \mgii/\feii
selected sample, or 0.02 for the 30 per cent with the strongest
absorption lines.  After correcting for dust obscuration bias, the
distribution of dust contents in the \caii absorbers starts to
approach that seen in emission-selected galaxies at similar redshifts,
a very different story to that seen in DLAs. Without \nhi measures it
is difficult to place constraints on their absolute metallicities,
however, their dust-to-metals ratio similar to that found in the Milky
Way suggests a high degree of chemical enrichment.  To summarise, the
absorption properties of the \caii absorbers point strongly towards
them tracing dense, metal and dust rich gaseous clouds. The number
densities of \caii absorbers, after correcting for dust obscuration
bias, is $\sim$30\,per cent that of DLAs at similar redshifts and a plausible
scenario was that, while DLAs could be found out to extended radii
from the center of galaxies, \caii absorbers traced the innermost,
star-forming, disks. 

In Wild et~al.  (2007) we showed that the average star formation within
1\farcs5--radii of strong \caii absorbers is four times that associated with
strong \mgii absorbers.  However, if it is assumed that the \caii absorbers
surround galaxies with close to unit filling factor, the galaxies responsible
for \caii absorption only account for $\sim$3\,per cent of the observed global
star formation\footnote{Modest corrections to obtain the true star formation
rate due to extinction by dust may be appropriate.}.

The results of the present paper reveal a very different picture for the nature
of the \caii absorber hosts.  We find there is a very strong bias towards
selecting luminous galaxies, and the \caii absorbing clouds are found out to
large physical distances from these luminous galaxies.  Our observations provide
an explanation for the low measured {\it in situ} star formation rate within
1\farcs5 ($\la$10\,kpc) of the \caii absorbers measured by Wild et~al.  (2007).
However, the origin of high column density, dusty, chemically enriched gas
clouds at up to 50\,kpc from luminous galaxies remains unclear.  There are
several competing theories and further observations are required to discriminate
between the models.
\begin{itemize}

\item Given the observed bias of our sample towards bright galaxies, a possible
explanation is that the \caii absorbers lie within extended gaseous disks such
as those being discovered with the GALEX satellite \citep{2005ApJ...619L..79T}.
These ultraviolet--bright disks are found to coincide with neutral hydrogen
structures, in which dust-to-gas ratios as high as the central regions of the
galaxy are found \citep{2003A&A...410L..21P}, in apparent agreement with the
relatively high dust contents of the \caii absorbers.

\item Possibly, the true absorber hosts are dwarf galaxies, associated, through
spatial clustering, with the larger galaxies seen in the $K$-band images.
Within the Local Group, the LMC and SMC lie at $\sim$50 and 60\,kpc from the
Milky Way, and while more than half of the MW satellites lie within 50\,kpc,
none contain significant quantities of gas \citep{2007ApJ...658..337B}.  A
similar distribution of satellites is seen for M31 \citep{2006MNRAS.365.1263M}.
Outside of our immediate neighbourhood, \citet{1993ApJ...405..464Z} find on
average less than 1.5 satellites\footnote{Satellites were defined as objects
2.2\,mag fainter than the primary galaxy;  most detected satellites were found
to be 2.2 to 6 magnitudes fainter than the primary.} within 500\,kpc of local
spiral galaxies.  Although the picture may be very different at $z\sim1$, the
10\,per cent covering factor for \caii absorption derived in the present paper
currently makes this scenario seem unlikely.  High resolution ultraviolet, or
possibly H$\alpha$, imaging (for example with the Hubble Space Telescope) would
reveal the existence of compact regions of star formation close to the absorber,
such as seen in the LMC, and distinguish between an origin for the absorbers in
the form of an extended disk associated with the luminous galaxies or from
discrete dwarf galaxies.

\item Perhaps \caii absorbers are the relics of outflows from previous
episodes of intense star formation in the luminous galaxies, but it is
not clear how dense clouds would survive the $\sim$10$^8$\,yr to reach
such distances at typical starburst outflow speeds without being
dispersed and the dust grains destroyed. \caii is regularly detected
in local High and Intermediate Velocity Clouds (HVCs and IVCs),
potential candidates for such outflow relics from the Milky
Way. However, none are found to have \caii column densities approaching
$1.7\times 10^{12}$ to $1.3\times10^{13}$cm$^{-2}$ as seen in our
systems \citep{2001ApJS..136..463W, 2006ApJ...638L..97T} (although
this may simply reflect lower Hydrogen column densities of the
clouds).  High-resolution imaging and spectroscopy of the galaxies
could indicate the likelihood of such a scenario by measuring the
relative angle of the major axis of the galaxy to the absorber and
determining the recent star formation history of the galaxies.

\item Finally, within standard galaxy evolution scenarios, tidal debris from
cannibalism of small galaxies by larger galaxies is expected to fill a much
greater volume around galaxies at $z$$\sim$1 than around galaxies today.  In the
local Universe, the SDSS has had a large impact on the detection of tidal
streams around the Milky Way.  Such streams cover hundreds of square degrees on
the sky and lie at several tens of kpcs from the center of the Milky Way
\citep{1997AJ....113..634I, 2003ApJ...588..824Y, 2006ApJ...642L.137B}.
Similarly to HVCs, the \hi column densities of the streams tend to be lower than
the DLA limit of $3\times10^{20}$:  mean columns in the Magellanic Stream are of
order $4\times10^{18}$ to $4\times10^{19}$
\citep{2003ApJ...586..170P}\footnote{The resolution of these measurements is
15.5\,arcmins.}.  \caii has been observed in absorption in the Magellanic Bridge,
but again with much lower column densities than found in our high redshift \caii
absorbers \citep{2005A&A...443..525S}.  Perhaps, a better local comparison might
be with the M81 group, in which a complex filamentary structure of high column
density neutral Hydrogen tails appear to have been caused by the interaction
between M81 and two neighbouring galaxies \citep{1994Natur.372..530Y}.  NGC 5291
\citep{2007astro.ph..3002B} is another example of a system where significant
columns of gas and associated star formation extend over large regions
surrounding prominent galaxies.

\end{itemize}

The high equivalent width \caii absorption systems that comprise our
sample are extremely rare in the local Universe but an understanding
of the origins of the few that do exist may reveal the physical
mechanism responsible for strong \caii absorption.  Spectroscopy of a
small number of luminous galaxies associated with low-redshift \caii
absorbers has been undertaken by Zych et al.  (2007).  The galaxies,
predominantly spirals, possess high metallicities and exhibit
substantial star formation rates.  Further progress with such
investigations would greatly aid in our interpretation of results at
higher redshift.

\section*{acknowledgments}

We would like to thank Emma Ryan-Weber, Chris Thom, Martin Zwaan and
Alan McConnachie for valuable discussions, and the anonymous referee
for useful comments.  VW is supported by the MAGPOP Marie Curie EU
Research and Training Network.  The United Kingdom Infrared Telescope
is operated by the Joint Astronomy Centre on behalf of the U.K.
Particle Physics and Astronomy Research Council. An anonymous referee
provided a careful reading of the original manuscript that resulted in
a number of improvements to the paper.

Funding for the Sloan Digital Sky Survey (SDSS) has been provided by
the Alfred P. Sloan Foundation, the Participating Institutions, the
National Aeronautics and Space Administration, the National Science
Foundation, the U.S. Department of Energy, the Japanese
Monbukagakusho, and the Max Planck Society. The SDSS Web site is
http://www.sdss.org/.  The SDSS is managed by the Astrophysical
Research Consortium (ARC) for the Participating Institutions. The
Participating Institutions are The University of Chicago, Fermilab,
the Institute for Advanced Study, the Japan Participation Group, The
Johns Hopkins University, Los Alamos National Laboratory, the
Max-Planck-Institute for Astronomy (MPIA), the Max-Planck-Institute
for Astrophysics (MPA), New Mexico State University, University of
Pittsburgh, Princeton University, the United States Naval Observatory,
and the University of Washington.

\bibliographystyle{mn2e}

\begin{appendix}\label{sec:app}
\section{Notes on Individual Objects}

The general discussion in Section \ref{sec:psfsub} and the image-subtractions
shown in Fig. 1 summarises the status of the subtractions for the majority of
objects. However, in a small number of cases, some specific comments are
warranted:

\vspace{0.25cm}
\noindent SDSS J085556.63+383232.7 ($z_{abs}$=0.852; $z_{quasar}$=2.065):  the
low surface brightness structure to the north-west of the quasar does not
reproduce consistently when using different PSF-stars in the subtractions.

\vspace{0.2cm}
\noindent SDSS J093738.04+562838.8 ($z_{abs}$=0.980; $z_{quasar}$=1.798):  the faint
apparent `donut' residual does reproduce when different PSF-stars are used.
However, the residual is slightly off-centre from the quasar image and extensive
interactive scaling of the PSF-stars fails to produce a convincing galaxy-like
image.  The quasar redshift is high, making the identification of a host galaxy
as the origin of the residual unlikely but it is possible that a galaxy
with a small galaxy-quasar separation, $\Delta\theta \la 0\farcs3$, is present.

\vspace{0.2cm} 
\noindent SDSS J105930.50+120532.8 ($z_{abs}$=1.050; $z_{quasar}$=1.570):  the
subtraction shown employs a rather faint PSF-star from the quasar image itself.
The noise level is thus enhanced but there is no systematic flux excess above
the the $K=20.0$ detection threshold for the galaxy catalogue.  PSF-stars from
the library of PSF-stars confirm the lack of any extended residual but at the
expense of more prominent `cloverleaf' residuals on small scales.

\vspace{0.2cm}
\noindent SDSS J110729.03+004811.2 ($z_{abs}$=0.740; $z_{quasar}$=1.391):
the subtraction shown employs a PSF derived from seven relatively faint stars
in the frame. The noise level in the subtracted frame is enhanced but the 
{\sc SExtractor} image analysis of the PSF-subtracted frame results in no
object brighter than $K=20.3$ being detected. Individual PSF stars from the 
PSF-library result in small-scale `butterfly' residuals but provide no
evidence for the presence of any galaxy image with $K\la 20.5$

\vspace{0.2cm}
\noindent SDSS J122756.35+425632.4 ($z_{abs}$=1.045; $z_{quasar}$=1.310): the faint
apparent residual approximately 0\farcs5 to the south does not reproduce well
when different PSF-stars are used.

\end{appendix}

\end{document}